\documentclass[letterpaper]{JHEP3}

\usepackage{epsfig}
\usepackage{amsmath}
\usepackage{xspace}
\newcommand{\HH}{\mathcal{H}}
\newcommand{\pd}[2]{\frac{\partial #1}{\partial #2}}

\title{ Ricci flow and black holes }

\author{Matthew Headrick \\
Center for Theoretical Physics, Massachusetts Institute of Technology \\ 77 Massachusetts Ave., Cambridge MA 02139, USA \\
E-mail: \email{headrick@mit.edu}
}

\author{Toby Wiseman \\
Jefferson Physical Laboratory, Harvard University \\
Cambridge MA 02138, USA \\
E-mail: \email{twiseman@fas.harvard.edu}
}

\date{June 2006}

\preprint{\hepth{0606086} \\ HUTP-06/A0022 \\ MIT-CTP/3754}

\abstract{
Gradient flow in a potential energy (or Euclidean action) landscape
provides a natural set of paths connecting different saddle points. We
apply this method to General Relativity, where gradient flow is Ricci
flow, and focus on the example of 4-dimensional Euclidean gravity with
boundary $S^1\times S^2$, representing the canonical ensemble for
gravity in a box. At high temperature the action has three saddle
points: hot flat space and a large and small black hole. Adding a time
direction, these also give static 5-dimensional Kaluza-Klein
solutions, whose potential energy equals the 4-dimensional action.
The small black hole has a Gross-Perry-Yaffe-type negative mode, and
is therefore unstable under Ricci flow. We numerically simulate the
two flows seeded by this mode, finding that they lead to the large
black hole and to hot flat space respectively, in the latter case via
a topology-changing singularity. In the context of string theory these
flows are world-sheet renormalization group trajectories. We also use
them to construct a novel free energy diagram for the canonical
ensemble.

}
\begin{document}
%

%
\section{Introduction and summary}
\label{sec:intro}
%

Gradient flow is a powerful theoretical tool for exploring a system's potential energy landscape, which has been applied in contexts ranging from biophysics to string theory. Similarly, it is sometimes useful to study gradient flow of a system's \emph{action} on its space of histories, for example in the context of the path integral formulation of quantum mechanics. (Like the path integral, the gradient flow will strictly speaking be well defined only in the Euclidean framework.)

The purpose of this paper is to investigate gradient flow, both of the potential energy and of the Euclidean action, in the context of General Relativity. Gradient flow of the Einstein-Hilbert action on the space of metrics on a given manifold is governed by the following equation:
\begin{equation}\label{RFone}
\pd{g_{\mu\nu}}{\lambda} = - 2 R_{\mu\nu},
\end{equation}
where $\lambda$ is the flow time. In the mathematical literature, the flow defined by equation \eqref{RFone} is known as Ricci flow, and in recent years has been the subject of intense study. It also arises in string theory as the one-loop approximation to the renormalization group (RG) flow of sigma models. Section 2 of this paper is devoted to a review of the basic properties of Ricci flow.

As our laboratory for the application of Ricci flow to gravity, we
take one of the simplest systems that admits multiple saddle points:
4-dimensional pure gravity (without cosmological constant) in a
spherical box of radius $R$, in the canonical ensemble at temperature
$\beta^{-1}$. The path integral for this system involves all Euclidean
metrics, on all topologies, with boundary $S^1_\beta\times S^2_R$. At
high temperature there are three saddle points of the action: a large
and a small black hole, and the product of $S^1_\beta$ with a flat
3-ball. The Lichnerowicz operators for the large black hole and hot
flat space metrics are positive definite, whereas, as we show, the one
for the small black hole has a single negative eigenvalue, analogous
to that of Gross, Perry, and Yaffe (GPY) \cite{GPY} in the
asymptotically flat case. This indicates that the small black hole is
truly a \emph{saddle point} of the action, whereas the other two
solutions are actually local minima (not counting Weyl fluctuations of
the metric, which always decrease the action; these are dealt with in
the path integral by rotating their contours of integration). GPY
argued that, due to this negative mode, the small black hole serves as
a bounce (or sphaleron) that allows the system to pass from one
minimum to the other by thermal fluctuations. Furthermore, according
to arguments given by Whiting and York \cite{WhitingYork}, Whiting
\cite{Whiting}, Prestidge \cite{Prestidge}, and Reall \cite{Reall},
this negative mode is related to the local thermodynamic instability
of the corresponding Lorentzian black hole.

We can also add a time direction to the system, and consider
5-dimensional Lorentzian gravity with boundary $\mathbf{R}\times
S^1_\beta\times S^2_R$. Ignoring the Hamiltonian constraint (which in
any case is automatically satisfied at the saddle points), the space
of time-symmetric initial data is the space of Euclidean 4-metrics,
and the potential energy is the 4-dimensional Euclidean action. The
three saddle points give static solutions, which are reinterpreted in
this context as large and small bubbles of nothing and the
Kaluza-Klein vacuum, respectively. Their dynamical stability is
controlled by the Lichnerowicz operators for the respective Euclidean
4-metrics, so the small bubble has a tachyon while the other two
solutions are stable. Further details about both the Euclidean and the
Lorentzian versions of this system are given in Section 3.

In both of the above contexts, Ricci flow describes gradient flow: of
the potential energy in the 5-dimensional Lorentzian context, and of
the action in the 4-dimensional Euclidean one. The small black hole's
negative mode indicates that it is unstable under Ricci flow. This
raises the question of where the two flows that start at this saddle
point, moving opposite ways along the unstable direction, ultimately
end up. One guess would be that each flow asymptotes to one of the
other two saddle points. Using numerical simulations we were able to
confirm this guess; this is the main result of this paper. Ricci flow
thus defines a unique path in the space of metrics connecting the
small black hole to each of the other two saddle points. The one
connecting the small black hole to hot flat space is particularly
interesting, because at a finite flow time the metric passes through a
localized topology-changing singularity (the metrics are otherwise
smooth throughout the flows). The flows are described in detail in
Section 4.

These flows have various applications. In the context of string theory, they represent world-sheet RG trajectories, with the small black hole as the ultraviolet fixed point and hot flat space and the large black hole as the infrared ones. It is interesting that they qualitatively mirror the time evolution of the Lorentzian small black hole perturbed by its thermodynamic instability, as it either evaporates or grows to become the large black hole. Quantitatively, however, the evolution is different, since at every point along the evaporation process the geometry remains Schwarzschild to a good approximation, whereas as we will see along the RG flows it does not.

RG flow is also often applied to the spatial part of a world-sheet
theory as a technique for studying closed-string tachyons. Hence we
may use the terminology of the 5-dimensional Lorentzian context and
say that the small bubble of nothing (which has a tachyon, and
therefore a relevant operator on the world-sheet) is the UV fixed
point, and the large bubble and Kaluza-Klein vacuum are the IR fixed
points. These flows are in accord with the arguments made in the paper
\cite{Gutperle:2002ki}, that the ADM energy of the target space should
be lower in the IR than in the UV of a world-sheet RG flow. The full
dynamical evolution of the unstable small bubble has been performed by
Lehner and Sarbach \cite{Lehner1,Lehner2}, and we may contrast
this behavior with the RG flows. In the case where the small bubble
initially expands, the dynamical simulations were consistent with the
large stable bubble being the end state, although the simulations were
not performed in a box and so saw only continued expansion. The more
interesting case is initial collapse of the small bubble, which in RG
flow goes to Kaluza-Klein flat space, but dynamically forms a
Kaluza-Klein black string. Whilst this will indeed evaporate to flat
space eventually, RG flow and classical dynamics clearly do not agree
for this system. We attribute the difference to the fact that this
flow involves a topology change, and whilst this may occur smoothly in
RG flow, in the dynamical case cosmic censorship is expected to form a
horizon to shield it.

Another application of our flows is in the Euclidean (thermal)
context. The fact that Ricci flow is gradient flow means that, at any
point along the flow, the action is stationary in the directions
orthogonal to it. Hence in principle these directions can be
integrated out to define an off-shell free energy; in a saddle point
approximation this free energy is simply equal to the value of the
action at that point on the curve. Thus, Ricci flow can be used to
construct a new type of free energy diagram for this thermal
system. This diagram, shown in figure \ref{fig:phase}, is discussed
further in Section 5.

Although we studied the specific case of 4 dimensions, it seems likely
that the flows are qualitatively similar in higher dimensions. Also,
while we employed a simple box as our infrared regulator, our results
would likely have been qualitatively unchanged had we instead
included a negative cosmological constant and imposed asymptotically
anti-de Sitter boundary conditions. It would be interesting to see
whether they have physical implications in the context of the AdS-CFT
correspondence, and we conclude the paper with a brief comment on
this.

Ricci flow has previously been studied numerically in a variety of contexts. For example, a 2-dimensional flow asymptoting to the dilaton black hole was studied by Hori and Kapustin \cite{Hori:2001ax}. The formation of singularities in 3 dimensions was studied by Garfinkle and Isenberg \cite{Garfinkle:2003an, MR2115754}. And Ricci flow was investigated as a numerical method for solving the Einstein equation on the 4-dimensional manifold K3 by the present authors \cite{HW}. However, as far as we know the flows we study in this paper are the first explicitly known that connect different saddle points of the action.

%
\section{Ricci flow}
\label{sec:RF}
%

Ricci flow is an analog of the heat equation for geometry, a diffusive process  acting on the metric of a Riemannian manifold. In this section we define it, explain its basic mathematical properties, and describe how it appears in several different contexts in physics. Indeed, historically it was in physics that Ricci flow first made its appearance, in the work of Friedan \cite{Friedan:1980jm} on the renormalization group for two-dimensional sigma models. It was subsequently introduced into mathematics by Hamilton \cite{MR664497} as a tool for studying the topology of manifolds. Among other things, he developed a program, recently brought to fruition by Perelman \cite{Perelman1,Perelman2}, for proving Thurston's geometrization conjecture concerning the classification of 3-manifolds (for reviews see \cite{MR2115067,Topping}). By now there is a very extensive literature (including a textbook \cite{MR2061425}) on Ricci flow.

The material in this section is review, except the discussion of boundary conditions in subsection 2.2, which is new as far as we know.

\subsection{Definition and basic properties}

Let $M$ be a $D$-dimensional compact manifold, and $\HH$ the space of Euclidean (i.e.\ positive definite) metrics $g_{\mu\nu}(x)$ on it (\emph{not} the space of metrics modulo diffeomorphisms). Ricci flow is the first-order flow on $\HH$, with respect to an auxiliary ``time" variable $\lambda$, defined by
\begin{equation}\label{RF}
\pd{g_{\mu\nu}(x,\lambda)}\lambda = - 2 R_{\mu\nu}(x,\lambda).
\end{equation}
The flow is obviously invariant under ($\lambda$-independent)
diffeomorphisms of $M$. It is sometimes useful to consider the more general equation
\begin{equation}\label{diffeoRF}
\pd{g_{\mu\nu}}\lambda = - 2 R_{\mu\nu} + 2\nabla_{(\mu}\xi_{\nu)},
\end{equation}
where $\xi^\nu(x,\lambda)$ is some vector field. In fact, it's easy to see that the solutions to equations \eqref{RF} and \eqref{diffeoRF} are the same, up to the diffeomorphism obtained by integrating $\xi^\mu$ with respect to $\lambda$. Thus the two equations describe different flows in $\HH$, but the same flow in the quotient of $\HH$ by diffeomorphisms of $M$.

In what sense is Ricci flow diffusive? Let us examine how a short-wavelength, small-amplitude perturbation of the metric evolves. If the wavelength is much shorter than the typical curvature radius of the manifold, then it is sufficient to expand about flat space. With $g_{\mu\nu} = \delta_{\mu\nu}+h_{\mu\nu}$, we find, to linear order,
\begin{equation}\label{flatperturb}
\pd{h_{\mu\nu}}\lambda = \nabla^2h_{\mu\nu} + 2\partial_{(\mu}v_{\nu)} + O(h^2),
\end{equation}
where
\begin{equation}
v_\nu \equiv 
\frac12\partial_\nu h_{\lambda\lambda} - \partial_\lambda h_{\lambda\nu}.
\end{equation}
We see that the perturbation evolves according to the heat equation, accompanied by a small diffeomorphism.\footnote{For pure gauge perturbations, which are of the form $h_{\mu\nu}=\partial_{(\mu}w_{\nu)}$ for some vector field $w_\nu$, the diffeomorphism term cancels the Laplacian term, since the Ricci tensor remains zero. For some purposes (for example in order to rigorously prove the short-time existence of solutions, or for improved numerical stability in simulating the flow) it is useful to have a strictly parabolic flow equation, in other words one where all short-wavelength modes decay. An elegant way to do this, due to DeTurck \cite{MR697987} and closely related to the harmonic gauge condition in General Relativity, is to fix an arbitrary connection $\tilde\Gamma^\mu_{\lambda\nu}$ on $M$ (thereby breaking the diffeomorphism invariance), and define the vector field
\begin{equation}\label{DTxi}
\xi^\mu = 
g^{\lambda\nu}(\Gamma^\mu_{\lambda\nu} - \tilde\Gamma^\mu_{\lambda\nu}),
\end{equation}
where $\Gamma^\mu_{\lambda\nu}$ is the connection compatible with $g_{\mu\nu}$. One then adds $2\nabla_{(\mu}\xi_{\nu)}$ to the flow equation, which only changes the flow by a $\lambda$-dependent diffeomorphism. Expanding $2\nabla_{(\mu}\xi_{\nu)}$ about flat space yields precisely $-2\partial_{(\mu}v_{\nu)}$, so that all short-wavelength perturbations evolve by the heat equation. As explained in Appendix A, a similar trick was employed in the numerical simulations of Ricci flow described in this paper.} Just as with the scalar heat equation, we are restricted to Euclidean signature, since on a Lorentzian background the flow is infinitely unstable to modes with timelike momentum, and therefore ill defined.

How unique is Ricci flow? Suppose that we wish to define a flow equation for the metric that is local and diffeomorphism-invariant and that contains at most two second derivatives acting on the metric. There are only two terms we could add to the right-hand side of \eqref{RF}:
\begin{equation}\label{generalRF}
\pd{g_{\mu\nu}}\lambda = - 2 R_{\mu\nu} + \alpha Rg_{\mu\nu} +2\Lambda g_{\mu\nu}.
\end{equation}
Because it contains no derivatives, the $\Lambda$ term does not change
the behavior of short-wavelength perturbations, and therefore does not
affect the locally diffusive nature of the flow. While it is useful
when one is interested in flows that start or end on Einstein metrics
with non-zero cosmological constant, in this paper we will set
$\Lambda=0$ for simplicity. On the other hand, having a non-zero
$\alpha$ changes the right-hand side of \eqref{flatperturb}, adding
the terms $\alpha(\partial_\lambda\partial_\rho h_{\lambda\rho} -
\nabla^2h_{\lambda\lambda})\delta_{\mu\nu}$, which has the effect of
making different polarizations diffuse at different rates. A given
perturbation can be decomposed into a part obeying
$\partial_\lambda\partial_\rho h_{\lambda\rho} -
\nabla^2h_{\lambda\lambda}=0$, which evolves independently of
$\alpha$, and a Weyl part (proportional to $\delta_{\mu\nu}$), which
evolves with a diffusion constant $1-\alpha(D-1)$. In particular,
unless $\alpha\le1/(D-1)$ the flow is ill defined. In this sense Ricci
flow is singled out; as we will discuss below, the value $\alpha=0$ is
also physically selected in the context of RG flow. It would be
interesting to understand how general our results are for the other
values of $\alpha$ for which the flow is well defined.

Now let us review the basic global properties of Ricci flow. The fixed points are obviously Ricci-flat metrics. The generalization of the linearization \eqref{flatperturb} to the case $g_{\mu\nu}=\hat g_{\mu\nu} + h_{\mu\nu}$, where $\hat g_{\mu\nu}$ is Ricci-flat, is
\begin{equation}\label{genperturb}
\pd{h_{\mu\nu}}\lambda = 
-\Delta_{\rm L}h_{\mu\nu} + 2 \nabla_{(\mu}v_{\nu)} + O(h^2),
\end{equation}
where $v_\nu \equiv \frac12\partial_\nu{h^\lambda}_\lambda - \nabla_\lambda{h^\lambda}_\nu$, and $\Delta_{\rm L}$ is the Lichnerowicz operator:
\begin{equation}\label{eq:stability}
\Delta_{\rm L}h_{\mu\nu} \equiv -\nabla^\lambda\nabla_\lambda h_{\mu\nu} - 2 {{{R_\mu}^\kappa}_\nu}^\lambda h_{\kappa\lambda}
\end{equation}
(everything is calculated with respect to the background metric $\hat
g_{\mu\nu}$). The linear stability of the fixed point is thus
determined by the existence of negative eigenvalues of $\Delta_{\rm
L}$. ($\Delta_{\rm L}$ is non-negative on pure gauge modes, so any
negative mode must be physical.) $\Delta_{\rm L}$ acts separately on
Weyl ($h_{\mu\nu}(x)=\epsilon(x)\hat g_{\mu\nu}(x)$) and traceless
(${h^\mu}_\mu(x)=0$) modes. For Weyl modes it reduces to minus the
scalar Laplacian, which is non-negative on a compact manifold. On the
other hand, for traceless modes the Riemann tensor acts as a
``potential" on the manifold (in the sense of the Schr\"odinger
equation), with regions of positive curvature corresponding to
negative potential energy. The operator may therefore have a finite
number of ``bound states", i.e.\ modes with negative eigenvalue.

Another important property of Ricci flow, obvious from the definition, is that regions (and, within a region, directions) with positive Ricci curvature tend to collapse. This can lead to the formation of curvature singularities in finite flow time. The simplest example is to start with the round metric on the $D$-sphere. We will see a more complicated example in the flows we study in this paper.

\subsection{Boundaries}

In Section 4 we will consider Ricci flow on a manifold with
boundary. We should therefore consider what boundary conditions are
required to render the flow well defined. Let us locally divide the
coordinates into ones that are parallel ($x^i$) and normal ($x^n$) to
the boundary. In general for a diffusive flow one should impose a
single boundary condition (such as Dirichlet or Neumann) on any
variable $f$ which is acted on by a second derivative normal to the
boundary, i.e.\ where $\partial_n^2f$ appears, while for any variable
acted on by fewer than two normal derivatives, no boundary condition
should be imposed. In the Ricci tensor only the metric components
along the boundary, $g_{ij}$, appear with two normal derivatives, so
it is only on these that we should impose boundary
conditions.\footnote{As explained in the previous subsection, the DeTurck 
flow is equivalent to Ricci flow up to a diffeomorphism but is
strictly parabolic, so that all metric components appear with second
derivatives in all directions. For this flow we will therefore need to
impose boundary conditions on all components of the metric. The extra
$D$ boundary conditions (on $g_{in}$ and $g_{nn}$) are provided by the
condition that $\xi^\mu$ (as defined in \eqref{DTxi}) vanishes on the
boundary. This is necessary for DeTurck flow to be equivalent to
Ricci flow with the prescribed boundary conditions, which requires
that the $\lambda$-dependent diffeomorphism that relates them be the
identity on the boundary.} For example, in the Dirichlet case we would
fix the induced metric $\gamma_{ij}$ on the boundary, while in the
Neumann case we would fix its extrinsic curvature $K_{ij}$ (the latter
case is treated in the paper \cite{MR1387799}). These are the same
kinds of boundary conditions one can impose on the hyperbolic time
evolution of general relativity.

In what follows, we will adopt Dirichlet boundary conditions. $\HH$ is therefore defined as the space of Euclidean metrics on $M$ with a fixed induced metric $\gamma_{ij}$ on $\partial M$. An important technical point is that, with these boundary conditions, we cannot make the usual decomposition of perturbations into Weyl and traceless parts. Specifically, a perturbation whose component $h_{nn}$ doesn't vanish at the boundary cannot be expressed as the sum of an allowed Weyl and an allowed traceless perturbation.

\subsection{Ricci flow as gradient flow}

The fixed points of Ricci flow, namely the Ricci-flat metrics, coincide for $D>2$ with the critical points of the Euclidean Einstein-Hilbert action,
\begin{equation}\label{eq:action}
S[g_{\mu\nu}(x)] = - \frac{1}{16 \pi G_{\rm N}} \int_{M}
\sqrt{g} R - \frac{1}{8 \pi G_{\rm N}} \int_{\partial M} \sqrt{\gamma} K + S_0
\end{equation}
($S_0$ is an arbitrary constant). This suggests that Ricci flow might be gradient flow of $S$. We will now show that this is the case.

Recall that the general gradient flow equation, on a space with coordinates $g^A$ and ``energy" function $S$, is
\begin{equation}
\frac{dg^A(\lambda)}{d\lambda} = -G^{AB}\pd S{g^B}.
\end{equation}
The inverse metric $G^{AB}$ is necessary to raise the index on the gradient. In our case $g^A$ is $g_{\mu\nu}(x)$, the index $A$ standing for both the point $x$ in $M$ and the component $\mu\nu$, and $S$ is the Einstein-Hilbert action. But what should we take for $G_{AB}$? The most general metric on $\HH$ that is local and diffeomorphism-invariant (on $M$) and contains no derivatives is of the form
\begin{equation}\label{genHmetric}
G^{(a)}_{AB}dg^Adg^B = 
\frac1{32 \pi G_{\rm N}}\int_M\sqrt{g}
\left({dg^\mu}_\nu{dg^\nu}_\mu + a({dg^\mu}_\mu)^2\right),
\end{equation}
where ${dg^\mu}_\nu=g^{\mu\lambda}dg_{\lambda\nu}$. The only free parameter is $a$ (except the trivial overall factor which has been fixed for convenience). With this metric we find
\begin{equation}
-G^{AB}_{(a)}\pd S{g^B} = -2R_{\mu\nu}+\alpha Rg_{\mu\nu},
\end{equation}
where
\begin{equation}
\alpha = \frac{2a+1}{Da+1}.
\end{equation}
Gradient flow thus establishes a one-to-one map between $a$ and the parameter $\alpha$ from \eqref{generalRF}. In particular, Ricci flow ($\alpha=0$) is obtained when $a=-1/2$. We will simply refer to this metric as $G_{AB}$:\footnote{In a previous version of this paper, we mistakenly referred to this metric as the ``DeWitt" metric. In fact, in the DeWitt metric, which plays an important role in the Hamiltonian formulation of General Relativity \cite{DeWitt}, $a$ takes the value $-1$. We would are very grateful to S. Hartnoll for pointing this out, as well as for several other very useful comments on this subsection of the paper.}
\begin{equation}\label{Hmetric}
G_{AB}dg^Adg^B = 
\frac1{32 \pi G_{\rm N}}\int_M\sqrt{g}
\left({dg^\mu}_\nu{dg^\nu}_\mu - \frac12({dg^\mu}_\mu)^2\right).
\end{equation}

Notice that $G_{AB}$ is not positive definite, as Weyl perturbations
have negative norm. This is somewhat unconventional for a gradient
flow, and implies that $S$ does not necessarily decrease along the
flow (hence we should avoid the terminology ``gradient descent"). The
fact that $G_{AB}$ is negative on Weyl modes is actually crucial for
obtaining a well-defined gradient flow from the Euclidean
Einstein-Hilbert action. (In fact, notice that more generally the
condition $\alpha<1/(D-1)$ mentioned below \eqref{generalRF} requires
$a<-1/D$, i.e.\ requires the Weyl modes to have negative norm.) The
reason is that short-wavelength Weyl modes \emph{decrease} the value
of $S$. A conventional gradient descent would be therefore be unstable
to all such modes. However, if they have negative norm then the flow
runs \emph{uphill} in those directions, rendering it well behaved.

The Weyl modes' negative norm plays a similar role in Euclidean
Quantum Gravity, where one integrates $e^{-S}$ over $\HH$. In that
context, the fact that these modes decrease the action, which is
therefore not bounded below, threatens to make the path integral ill
defined. The standard cure is to rotate the contour of integration for
each Weyl mode to lie parallel to the imaginary axis \cite{GHP}.  In
analogy with the path integral for a particle or string moving in a
Lorentzian background, where the contours for the timelike target
space direction are similarly rotated, this prescription is quite
natural if one takes $G_{AB}$ as the metric on $\HH$. We know of no
argument for Euclidean Quantum Gravity that requires this metric to
have $a = -1/2$. This value naturally gives rise to the de Donder gauge condition when removing pure gauge modes from the one-loop determinant, as we will review in subsection 3.1. 
However, it would be interesting to study the path integral and its gradient
flows arising from different choices of $a$.

\subsection{Relation to sigma-model RG flow}

A two-dimensional sigma model is defined, at a given energy scale
$\Lambda$, by a target space manifold $M$ equipped with a metric
$g_{\mu\nu}(x)$, a two-form $B_{\mu\nu}(x)$, and a dilaton
$\Phi(x)$. Let us set $B_{\mu\nu}=0$ (which is equivalent to imposing
parity symmetry). To first order in the coupling $\alpha'$, the
running of the metric and dilaton is given by
\begin{equation}\label{RG}
\pd{g_{\mu\nu}}\lambda = -2 R_{\mu\nu} -4\nabla_\mu\partial_\nu\Phi, \qquad
\pd\Phi\lambda = -\frac D{3\alpha'} + \nabla^2\Phi - 2 (\partial_\mu\Phi)^2,
\end{equation}
where $\lambda = -\frac12\alpha'\ln\Lambda$. (We have written the equations
for the bosonic sigma model, but the only change for the
supersymmetric versions at this order in $\alpha'$ is the coefficient
of $1/\alpha'$ in the dilaton equation.) The flow equation for
$g_{\mu\nu}$ is of the form \eqref{diffeoRF}, with $\xi_\mu =
-2\partial_\mu\Phi$. Therefore, irrespective of how the dilaton
evolves, the metric evolves according to Ricci flow, up to a
$\lambda$-dependent diffeomorphism. (However, one as to be
careful: because of the dilaton a fixed point of RG flow is not necessarily a fixed point of Ricci flow, and vice versa.) Alternatively, one can consistently set the dilaton to a constant, in which case $g_{\mu\nu}$ evolves strictly
according to Ricci flow.

Just as Ricci flow is gradient flow of the Einstein-Hilbert action, the RG flow defined by \eqref{RG} is gradient flow of the spacetime effective action of string theory,
\begin{equation}
S[g_{\mu\nu}(x),\Phi(x)] = 
-\frac1{2\kappa^2}\int_M\sqrt ge^{-2\Phi}\left(-\frac{2D}{3\alpha'} + R+4(\partial_\mu\Phi)^2\right)
-\frac1{\kappa^2}\int_{\partial M}\sqrt\gamma e^{-2\Phi}K,
\end{equation}
with respect to the metric
\begin{equation}
\frac1{4\kappa^2}\int\sqrt{g}e^{-2\Phi}\left({dg^\mu}_\nu {dg^\nu}_\mu-\frac12({dg^\mu}_\mu-4d\Phi)^2\right).
\end{equation}
Note that the timelike direction is now a particular mixture of the dilaton and the Weyl mode of the metric.

The RG flow of two-dimensional quantum field theories is a subject of
physical interest in its own right. Recently it has also become a
popular tool for exploring the off-shell configuration space of string
theory, for example in the study of tachyon condensation.\footnote{See
\cite{Freedman:2005wx} and references therein for a discussion of the
relationship between world-sheet RG flow and spacetime dynamics of
closed string tachyons.} We will not include a dilaton in the analysis
that follows, but since as explained above the dilaton doesn't enter
into the running of the metric, our work can be seen as an off-shell
exploration either of pure gravity or of string theory.

%
\section{Gravity in a cavity}
\label{sec:box}
%

In this paper we will study Ricci flow on 4-manifolds with boundary
$S^1\times S^2$, employing Dirichlet boundary conditions with the
boundary metric fixed to be a round sphere of radius $R$ times a
circle of circumference $\beta$. These boundary conditions arise in
two distinct physical contexts. The first is the canonical ensemble
for 4-dimensional gravity in a spherical box, with the circle
representing the imaginary time direction. The second is the
superspace for 5-dimensional gravity Kaluza-Klein reduced on a
circle, again in a spherical box.

We will begin by reviewing the canonical ensemble for gravity in a box. We will then explain how the mathematical features of this system are to be physically reinterpreted in the Kaluza-Klein context.

\subsection{Thermal interpretation}

\begin{figure}
\centerline{\epsfig{file=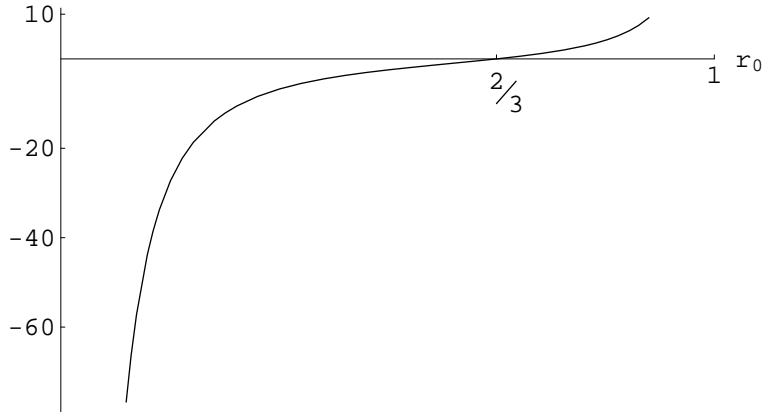,width=4in}}
\caption{ Lowest eigenvalue $\mu$ of the Lichnerowicz operator
$\Delta_{\rm L}$ on the black hole background (3.1). The small and
large black holes are separated by the value $r_0/R=\frac23$, where
the eigenvalue passes from negative to positive. At this point the
mode becomes tangent to this one-parameter family of solutions.  }
\label{fig:negmode}
\end{figure}

The canonical ensemble for gravity in asymptotically flat space at
finite temperature is ill defined due to infrared instabilities, both
perturbative (Jeans instability) and non-perturbative (nucleation of
black holes). An infrared regulator is therefore necessary, the
simplest one being the spherical box we employ here. Other regulators
are possible; for example one can introduce a negative cosmological
constant and work in asymptotically anti-de Sitter spaces. The
thermodynamics of the system doesn't depend qualitatively on which
regulator is used. What is described in this subsection was largely
the work of Gross, Perry, and Yaffe \cite{GPY}, Hawking and Page
\cite{HawkingPage}, and York \cite{York:1986it}, except for the
calculations of the negative mode for the small black hole (resulting
in figure \ref{fig:negmode}) which are new.

We consider pure gravity in a rigid spherical box of radius $R$, which
is kept in contact with a heat bath at a temperature
$\beta^{-1}$. Note that it is the \emph{boundary} that is kept at that
temperature; even though the system is in thermal equilibrium,
observers at different points inside the box may experience different
temperatures due to gravitational red- and blue-shifting. According to
the prescription of Euclidean Quantum Gravity, the partition function
for this system is obtained by integrating $e^{-S[g]}$ over all
Riemannian manifolds with boundary metric $S^1_\beta\times
S^2_R$. This involves both summing over topologies and integrating
over metrics.\footnote{In principle this includes disconnected
manifolds. However, in this case the given boundary is connected, so
any manifold satisfying the boundary conditions can be divided into
the connected component containing the boundary and the other
components. The other components factor out of the path integral,
multiplying the partition function by a constant factor. Therefore it
is sufficient to consider connected manifolds.}

The path integral is usually treated by the saddle point method. The
number of saddle points (i.e.\ Ricci-flat metrics) depends on the
dimensionless ratio $b \equiv \beta/(2\pi R)$. For any value of $b$,
we have hot flat space, $S^1_\beta\times B^3_R$, where $B^3_R$ is a
flat 3-ball of radius $R$. For $b<b_{\rm crit}=4/3\sqrt{3}\approx0.77$
(in other words, at sufficiently high temperature), there are two
other solutions. Both of them have topology $B^2\times S^2$, and both
have the Euclidean Schwarzschild metric, but with different values of
the horizon radius $r_0$:
\begin{equation}
ds^2 = 
4r_0^2\left(1 - \frac{r_0}{r}\right)d\tau^2 + \left(1 - \frac{r_0}{r}\right)^{-1} dr^2 + r^2 d\Omega_2^2.
\label{eq:metric}
\end{equation}
The radial coordinate $r$ goes from $r_0$ (at the horizon) to $R$ (at the boundary). The imaginary time coordinate has periodicity $2\pi$, and the factor of $4r_0^2$ in the metric ensures smoothness at the horizon. For the circle to have the right length at the boundary, we must have
\begin{equation}
\beta = 4 \pi r_0\sqrt{ 1 - \frac{r_0}{R} },
\label{eq:betaeq}
\end{equation}
which fixes $r_0$ in terms of $\beta$ and $R$. For $b<b_{\rm crit}$ there are two solutions, with $r_0$ greater than and less than $\frac23R$ respectively, which we refer to as the small and large black holes. Although it has not been proven, it is believed that these are the only saddle points for these boundary conditions \cite{gary}, and this is what we will assume. In the limit that we remove the cutoff ($R\to\infty$, $\beta$ fixed), the small black hole goes over to the usual asymptotically flat Schwarzschild black hole ($r_0\to\beta/4\pi$), while the large black hole disappears ($r_0\to\infty$). Hot flat space persists as well in this limit, of course.

\begin{figure}
\centerline{\epsfig{file=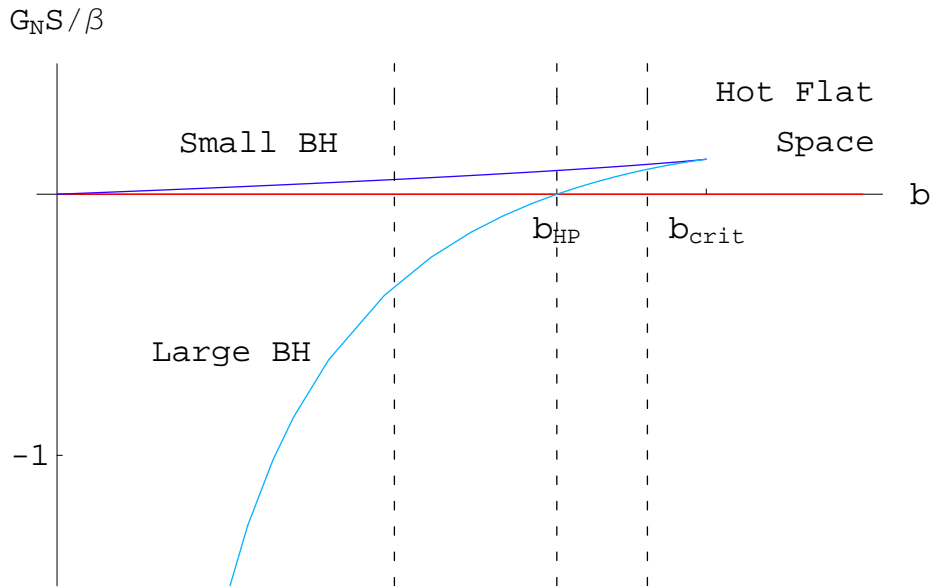,width=5in}}
\caption{Action versus inverse temperature parameter
$b\equiv\beta/(2\pi R)$ for the three saddle points. For $b>b_{\rm
crit}$ only hot flat space is allowed, while for $b<b_{\rm crit}$
there are also small and large black hole solutions. At $b_{\rm HP}$
there is a first-order phase transition, the Hawking-Page transition:
for $b>b_{\rm HP}$ flat space has the lowest action and therefore
dominates thermodynamically, while for $b<b_{\rm HP}$ the large black
hole dominates. The dashed lines indicate the three values of $b$
which will be used in later figures to display the qualitative
behavior of the simulated Ricci flows. For this and all other figures
we have chosen units such that $R=1$.
\label{fig:box}
}
\end{figure}

The system's partition function, and therefore its thermodynamical
properties, are dominated by the saddle point with the lowest value of
the action \eqref{eq:action}. Choosing the constant $S_0$ to conform
to the convention that flat space has zero action, the Schwarzschild
solutions have action
\begin{equation}
S = \frac{\beta R}{G_{\rm N}}\left[1 + \frac{4\pi r_0}\beta\left(\frac{3r_0}{4R}-1\right)\right].
\end{equation}
The actions of the three saddle points are plotted as a function of
$b$ in figure \ref{fig:box} (for this and all other figures we choose
units such that $R = 1$). We see that, for all temperatures
where it exists, the small black hole has the largest action of all
the saddle points. At low temperatures, specifically for $b>b_{\rm
HP}=16/27\approx0.59$, hot flat space dominates, while at higher
temperatures the large black hole dominates. The first-order phase
transition separating these two regimes is known as the Hawking-Page
transition, by analogy with the same phenomenon in AdS
\cite{HawkingPage}.

So far we have neglected the one-loop determinant in the saddle-point approximation to the path integral (not to mention the higher-loop corrections, which require an ultraviolet cutoff, such as string theory, to define). This correction is indeed negligible at small $\hbar$ (i.e.\ when the system is much larger than the Planck length), \emph{as long as all the eigenvalues are positive}; otherwise the saddle point approximation doesn't make sense. The operator in question, taking into account the rotation of the contours for the conformal modes, is
\begin{equation}\label{Delta}
{\Delta^A}_C = G^{AB}\frac{\partial^2S}{\partial g^B\partial g^C}.
\end{equation}
Acting on a perturbation $h_{\mu\nu}$, ${\Delta^A}_C$ yields precisely
(minus) the right-hand side of \eqref{genperturb} (since the latter is
just $\partial_C(-G^{AB}\partial_BS)h^C$, and of course
$\partial_BS=0$ at the saddle point). This operator has infinitely
many zero eigenvectors corresponding to pure gauge fluctuations, of
the form $h^{\rm gauge}_{\mu\nu} = \nabla_{(\mu}w_{\nu)}$, which must
be removed before the determinant is calculated. Since ${\Delta^A}_C$
is self-adjoint (with respect to $G_{AB}$) the physical eigenvectors
are orthogonal to the pure gauge ones, and therefore correspond to
fluctuations obeying the de Donder gauge condition
$\nabla_\mu{h^\mu}_\nu-\frac12\partial_\nu h=0$ (since $G_{AB}h_{\rm
gauge}^Ah^B = \int\sqrt g\nabla_{(\mu}w_{\nu)}(h^{\mu\nu}-\frac12h
g^{\mu\nu}) = -\int\sqrt gw_\nu(\nabla_\mu
h^{\mu\nu}-\frac12\partial^\nu h)$). Restricting to the orthogonal
complement of the gauge zero eigenvectors leaves these physical
fluctuations on which ${\Delta^A}_C$ is just the Lichnerowicz operator
$\Delta_{\rm L}$. (Equivalently, one can follow the Faddeev-Popov
procedure and add a gauge-fixing term and ghosts \cite{GHP}.)

On the hot flat space background, $\Delta_{\rm L}$ is positive since
the Riemann tensor vanishes and we have imposed Dirichlet boundary
conditions.\footnote{At extremely high temperatures there are more
subtle instabilities which we do not consider here. When $\beta$ is
smaller than about $G_{\rm N}^{1/4}R^{1/2}$ the gas of thermal
gravitons suffers from the Jeans instability \cite{GPY}. Within string
theory, when $\beta$ is smaller than the string scale there will be an
unstable winding state, indicating a Hagedorn phase transition
\cite{Atick}.}  What about on the black hole backgrounds? In figure
\ref{fig:negmode}, the lowest eigenvalue of $\Delta_{\rm L}$ is
plotted against $r_0$. For the large black hole ($r_0>\frac23R$) it is
positive, so like flat space it is indeed a good candidate for the
thermal ground state of the system. In the regime where each has the
lowest action, that action (divided by $\beta$) is the classical
approximation to the system's free energy. The small black hole, on
the other hand, has a negative eigenvalue mode, the cavity
generalization of the Gross-Perry-Yaffe instability of an
asymptotically flat black hole \cite{GPY}. As in the asymptotically
flat case, this negative mode is mainly localized near the horizon,
and preserves the $U(1) \times SO(3)$ isometries of the black hole
(i.e.\ only depends on the radial coordinate). In past literature
there is some confusion over the calculation of this mode in a cavity
\cite{Allen,York:1986it,Prestidge}. The important point is that,
unlike in the asymptotically flat case, the mode is not traceless,
since the cavity boundary conditions (constant Euclidean circle and
2-sphere radius) explicitly couple the trace and traceless modes. When
this is taken into account one finds the curve in figure
\ref{fig:negmode}. (See Appendix A for details of the calculation.)
In particular, as expected, one finds that the instability disappears
exactly at the transition from the small to the large black hole, and
therefore at the transition from local thermodynamic instability to
stability. As pointed out by Hawking and Page \cite{HawkingPage}, the
zero mode that appears corresponds to slightly changing the black
hole's radius without altering the boundary condition.

What is the meaning of this negative mode? It has been argued to be
related to the fact that the corresponding Lorentzian solution is
locally thermodynamically unstable since it has a negative specific
heat \cite{WhitingYork,Prestidge,Reall}. The negative mode also means
that the Euclidean solution can serve as a bounce, representing
thermal fluctuations that take hot flat space to the large black hole
and vice versa, allowing the system to find its true ground state
whichever side of the Hawking-Page transition it's on
\cite{GPY,York:1986it}. So although it is never the lowest-action
saddle point, the small black hole nevertheless plays an important
physical role in the system's dynamics.

\subsection{Kaluza-Klein interpretation}

The three 4-dimensional Euclidean Ricci-flat geometries discussed in
the previous subsection have an entirely different physical meaning if
we interpret the circle direction, which was the imaginary time in the
thermal context, instead as a Kaluza-Klein circle, and add a real time
direction. 
The resulting 5-dimensional Lorentzian
geometries are static solutions to the 5-dimensional Einstein equation
with boundary $\mathbf{R}\times S^1_\beta\times S^2_R$.  Note that,
whereas in the 4-dimensional context these solutions are believed to
be the only ones, in the 5-dimensional context there are certainly
other static solutions, such as the uniform black string, which is (4-dimensional Lorentzian Schwarzschild)$\times S^1_\beta$. This is
because there is an additional degree of freedom present, namely the
time-time metric component.

What do these 5-dimensional solutions look like? First note that the
ADM energy of each is equal to the Euclidean action of the
corresponding 4-dimensional solution, plotted in figure
\ref{fig:box}. Furthermore, its linear stability is controlled by the
4-dimensional Lichnerowicz operator, since each mode of that operator
can be put on shell by dressing it with a time dependence
$e^{\pm i\sqrt\mu t}$, where $\mu$ is the $\Delta_{\rm L}$ eigenvalue. Let
us fix $\beta$. For all values of $R$ we have ``hot flat space", which
in this context is the Kaluza-Klein vacuum. As pointed out by Witten,
although it is perturbatively stable, in the limit $R\to\infty$ the
Kaluza-Klein vacuum is unstable due to nucleation of bubbles of
nothing \cite{Witten:1981gj}. At large but finite $R$ we might therefore expect to find a ground state with negative energy; indeed we
find the ``large black hole", which in this context is a static bubble
of nothing that almost fills the container, but leaves a shell of
``something" around the edge. This is also perturbatively stable. It
seems likely that, for $b>b_{\rm HP}$, the true ground state is the
Kaluza-Klein vacuum, while for $b<b_{\rm HP}$ it is the large bubble
of nothing. Finally, we have the ``small black hole", or small bubble
of nothing, which has higher energy than either the Kaluza-Klein
vacuum or the large bubble, and is unstable.\footnote{These solutions
are the analogs for zero cosmological constant of the asymptotically
AdS solutions found by Copsey and Horowitz \cite{Copsey:2006br}.} 

In the context of string theory, such a gravitational instability is
classified as a closed string tachyon. It is always interesting to
study the world-sheet RG flow seeded by the vertex operator of such a
tachyon, and to compare it to the system's dynamical evolution. The
classical evolution of the bubble perturbed by its unstable mode has
in fact already been studied by Sarbach and Lehner using numerical
simulation \cite{Lehner1,Lehner2}. Those simulations were performed
without the box, which simplifies the analysis because it allows
radiation to escape to infinity and the energy thereby to decrease. It
was found that, when perturbed in one direction, the bubble grew
without bound, while in the other it settled down to the black
string. Apparently cosmic censorship prevents the system from
classically reaching its local ground state, the Kaluza-Klein vacuum,
because that would entail the appearance of a naked topology-changing
singularity. (Of course quantum mechanically the black string would
presumably evaporate, eventually settling down to the vacuum.) We will
see in the next section that RG flow (which is well approximated by
Ricci flow when the system is much larger than the string scale) does
not suffer from such prudishness.

%
\section{Where does the small black hole flow?}
\label{sec:where}
%

Let us call the negative mode of the small black hole $h_{\mu\nu}^{\rm
GPY}$. If we perturb the small black hole by this mode and evolve the
metric by Ricci flow, then, as long as it remains sufficiently small,
the mode will grow exponentially. How does the metric evolve once it
leaves the linear regime? At late times does it converge to one of the
other saddle points? These are the questions we will answer in this
section. Note that there are two different flows to follow, depending
on whether $h_{\mu\nu}^{\rm GPY}$ is added with a positive or negative
coefficient. These are physically distinct; in one direction the mode
increases the horizon size and in the other decreases it. We will fix
the convention that $h_{\mu\nu}^{\rm GPY}$ increases the horizon
size. We performed numerical simulations of both of these Ricci flows
at several values of the parameter $b$, below, at, and above the
Hawking-Page value $b_{\rm HP}$. Details of the simulations are given
in Appendix A. In this section we summarize the main results.

First, however, we would like to point out two obvious generalizations of our work. The first is to dimensions $D>4$. The second is to the small black hole in anti-de Sitter space, in which case the flow equation must be modified to include the cosmological constant term:
\begin{equation}
\pd{g_{\mu\nu}}\lambda = - 2 R_{\mu\nu} + 2 \Lambda g_{\mu\nu}.
\end{equation}
In both cases the system is so closely analogous to the one we studied that it seems reasonable to guess that the behavior of the flows would be qualitatively the same. 

We begin with the flow seeded by the perturbation $+h_{\mu\nu}^{\rm
GPY}$. Since the perturbation does not break the black hole's
$U(1)\times SO(3)$ isometry group, the geometry enjoys this symmetry
throughout the flow. In principle we would consider an infinitesimal
perturbation. In practice for the initial data we perturbed the small
black hole such that the horizon was increased in radius by one
percent. This was sufficiently small that for some time the flow
remained well described by the linear theory.

\begin{figure}
\centerline{\epsfig{file=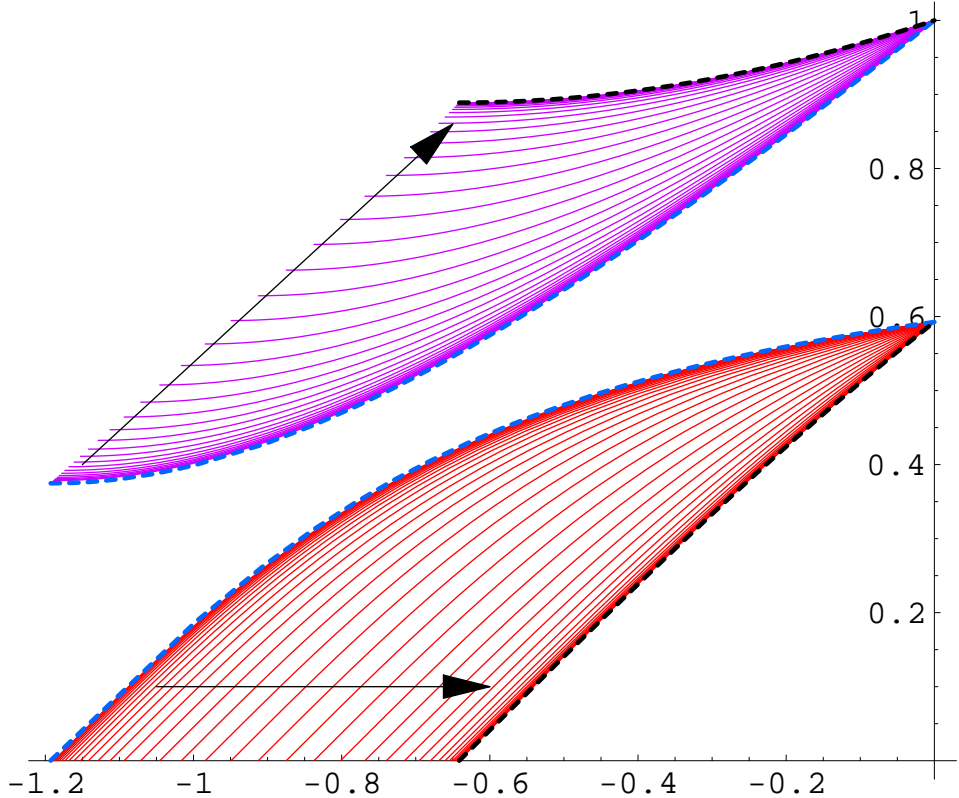,width=5.0in}}
\caption{Euclidean time radius $T$ (red) and sphere radius $S$
(purple) against proper radial coordinate $\rho$, for the flow of the
small black hole perturbed by $+h_{\mu\nu}^{\rm GPY}$, at $b = b_{\rm
HP}\approx0.59$. The boundary is at $\rho=0$. Snapshots are drawn at
intervals of $\lambda$ of $0.05$. The heavy blue curves show those
functions for the unperturbed small black hole. The heavy black curves
show them for the corresponding large black hole, to which the flow
clearly asymptotes at late times.
\label{fig:flowplus1}
}
\end{figure}

\begin{figure}
\centerline{
\epsfig{file=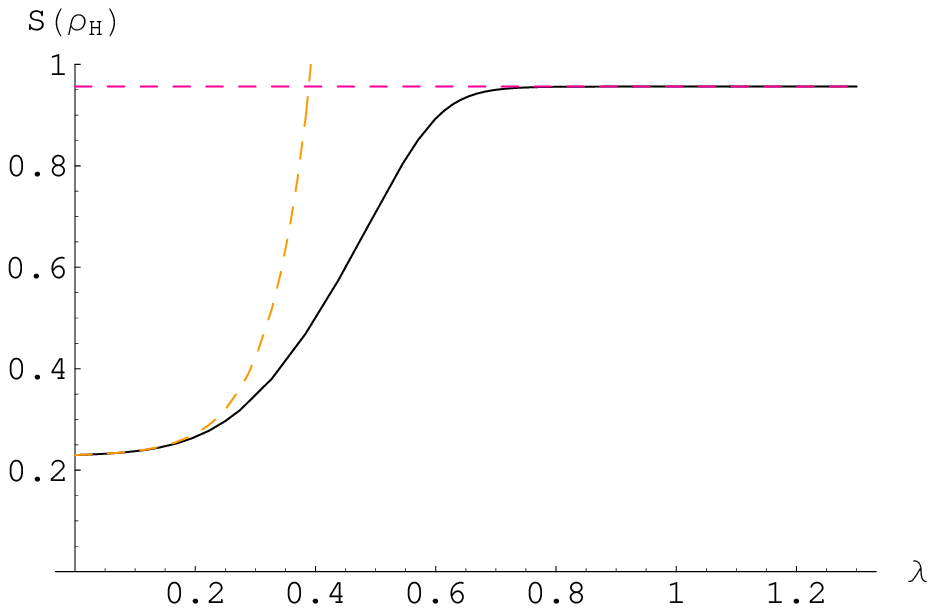,width=3.5in}
\epsfig{file=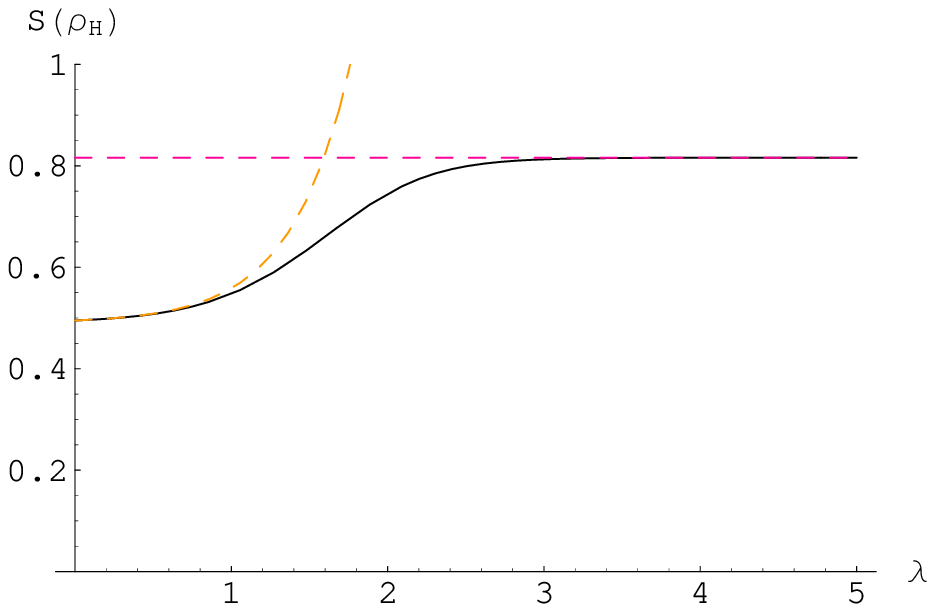,width=3.5in}
}
\centerline{
\epsfig{file=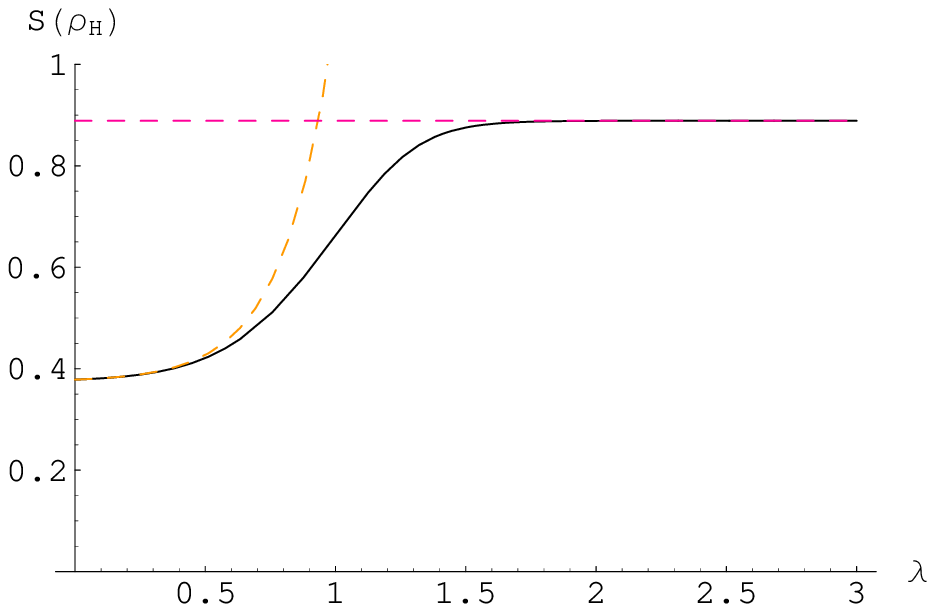,width=3.5in}
}
\caption{Horizon radius, $S(\rho_{\rm H})$, against $\lambda$ for
small black hole perturbed by $+h_{\mu\nu}^{\rm GPY}$, at $b = 0.40,
0.59, 0.70$ (the middle value equals $b_{\rm HP}$).  The horizontal
(red) dashed lines show the horizon radius for the corresponding large
black hole, which the metric asymptotically converges to in each
case. The curved (orange) dashed lines give the initial exponential
growth of the negative mode.
\label{fig:flowplus2}
}
\end{figure}

To characterize the geometry it is sufficient to know the radii of the
time circle and sphere, as functions of proper radial distance; in
other words, to eliminate the gauge ambiguity the metric may be put
into the form
\begin{equation}
ds^2 = T(\rho)^2 d\tau^2 + d\rho^2 + S(\rho)^2 d\Omega_2^2,
\label{eq:propermetric}
\end{equation}
setting $\rho = 0$ at the cavity wall (and $\rho < 0$ in the cavity
interior). $T(\rho)$ and $S(\rho)$ are plotted in figure
\ref{fig:flowplus1} for a representative value of $b$, namely $b_{\rm
HP}$. We see that the metric tends at large $\lambda$ to the large
black hole solution. Let $\rho_{\rm H}$ to be the value of $\rho$ at
the horizon. In figure \ref{fig:flowplus2} the horizon radius, i.e.\
$S(\rho_{\rm H})$, is plotted as a function of $\lambda$ for three
different values of $b$. Again, in each case it asymptotes to the
value for the large black hole. The initial growth of the small
perturbation is governed by the linear theory, growing exponentially
as $e^{-\mu \lambda}$, with $\mu$ given in figure
\ref{fig:negmode}. The same behavior was found for all values of $b$
tested: the metric relaxes to the large black hole solution.

\subsection{Singularity and surgery}

\begin{figure}
\centerline{\epsfig{file=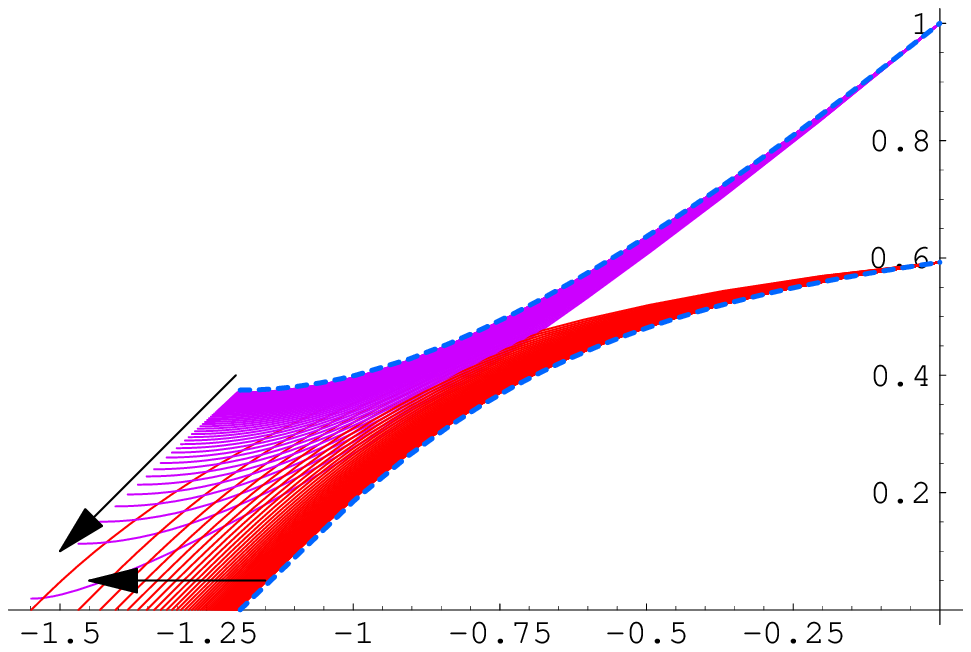,width=5.0in}}
\centerline{\epsfig{file=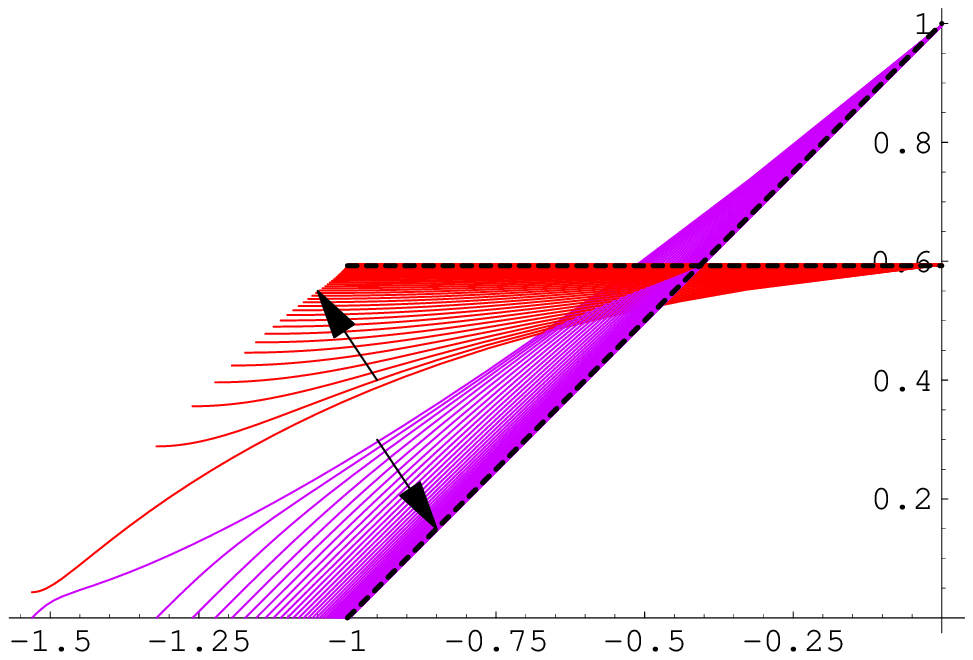,width=5.0in}}
\caption{
(Top) $T(\rho)$ (red) and $S(\rho)$ (purple) for the flow of the small black hole perturbed by $-h_{\mu\nu}^{\rm GPY}$, at $b = b_{\rm HP}$. Snapshots are drawn at intervals of $\lambda$ of $0.01$. The metric flows to a singularity, where the horizon shrinks to zero size. (Bottom) The metric after surgery, and the continuation of the Ricci flow. The metric asymptotically tends to flat space.
\label{fig:flowminus1}
}
\end{figure}

\begin{figure}
\centerline{\epsfig{file=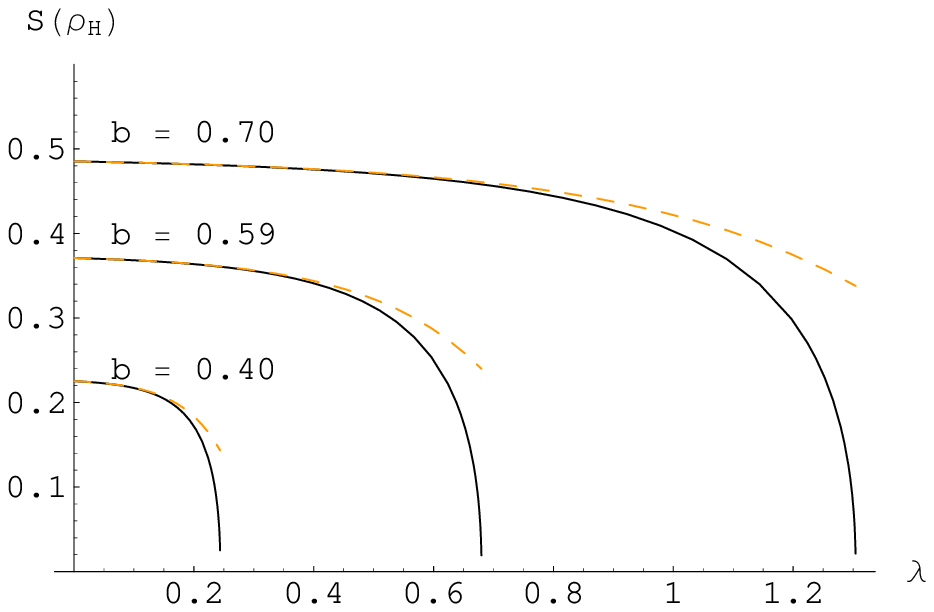,width=4in}}
\caption{Horizon radius, $S(\rho_H)$, as a function of $\lambda$ for
the flow of the small black hole perturbed by $-h_{\mu\nu}^{\rm GPY}$.
In all such flows the horizon collapses to
zero size in finite time. The dashed lines show the initial
exponential growth of the negative mode given by perturbation theory.
\label{fig:flowminus2}
}
\end{figure}

The behavior was quite different for the flows seeded by the
perturbation $-h_{\mu\nu}^{\rm GPY}$. For every value of $b$ tested,
after a finite flow time the horizon collapsed to a point, giving rise
to a curvature singularity and stopping the flow. An example is shown
in the top panel of figure \ref{fig:flowminus1}. In figure
\ref{fig:flowminus2}, the horizon radius is plotted against $\lambda$
for the same three values of $b$ as shown in figure
\ref{fig:flowplus2}. More details of the local model for the
singularity are given in appendix B.

\begin{figure}
\centerline{\epsfig{file=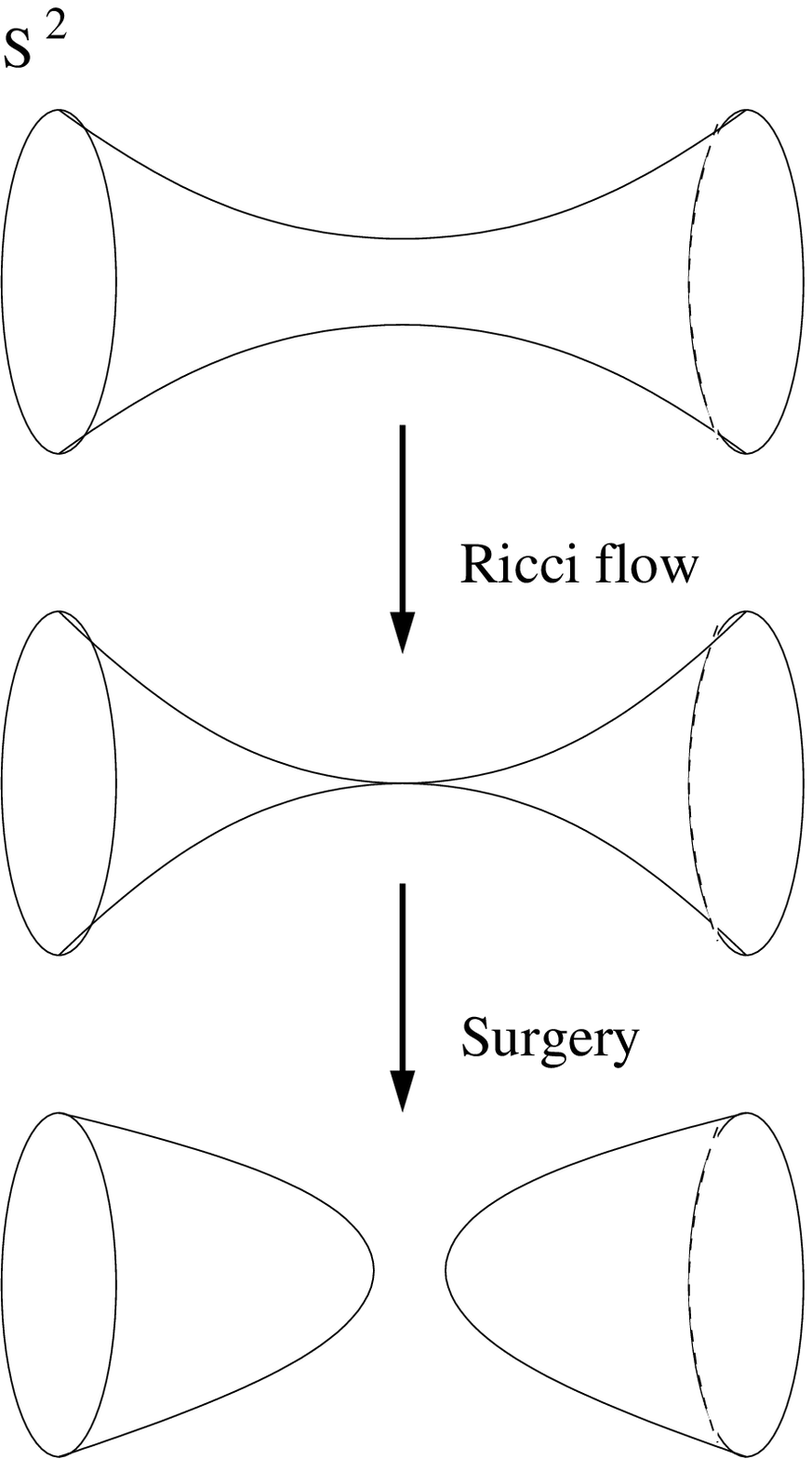,width=2.5in}}
\caption{An $S^2$ fibered over an interval will typically collapse under Ricci flow, producing a curvature singularity. This can be cured, allowing the flow to continue, by a local topology change known as a surgery.
\label{fig:surgery1}
}
\end{figure}

The spontaneous formation of singularities is generic in Ricci flow,
and has been much studied by mathematicians, particularly in the case
$D=3$. A common type of singularity that arises there, and is
analogous to what we observe here, involves an $S^2$ fibered with
varying radius over an interval (in a region of the manifold). The
fibers' positive curvature tends to make them shrink, and typically in
finite time one of them will collapse to zero size, giving rise to a
curvature singularity. In order to be able to continue the flow,
mathematicians perform ``surgery" on the singular manifold: a small
neighborhood of the singular point is excised and replaced with a
smooth metric. To avoid having the singularity re-appear, the topology
is changed. Specifically, a piece of the base is removed, and smooth
caps are sewn onto the two ends, as shown in figure
\ref{fig:surgery1}. As a result of this process, the manifold (or at
least this part of it) becomes disconnected, and the $S^2$ fibers,
which were previously non-contractible, become contractible.

\begin{figure}
\centerline{\epsfig{file=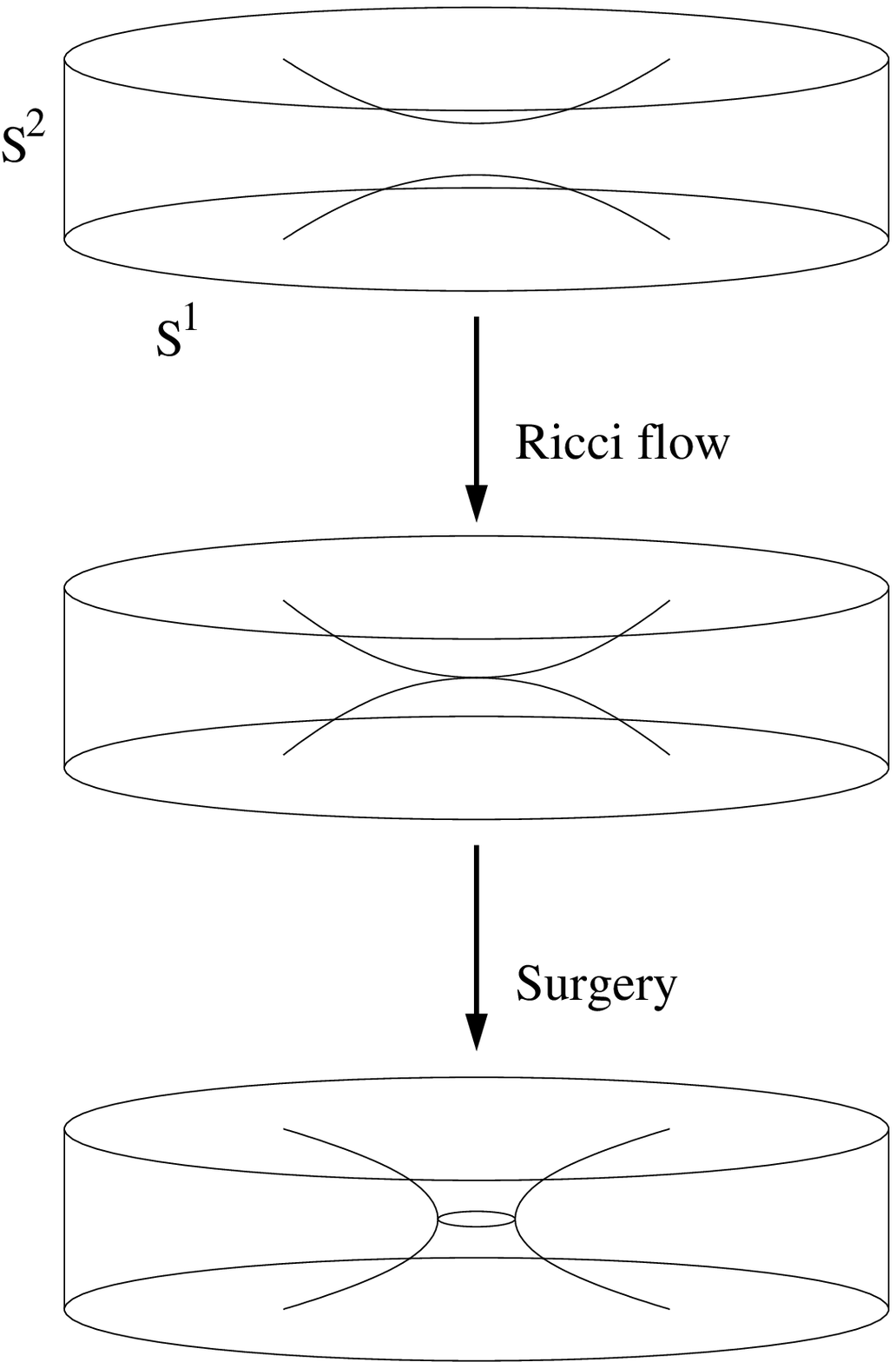,width=3in}}
\caption{When perturbed by $-h_{\mu\nu}^{\rm GPY}$, the small black
hole geometry (which is a 2-sphere fibered over a disk) evolves under
Ricci flow until the horizon (the fiber over the center of the disk)
collapses, giving rise to a singularity. The surgery we perform on the
singular manifold involves excising a small disk from the base, making
it an annulus, and making the sphere go to zero size on the inner
boundary so that the resulting total space is smooth and topologically
$S^1\times B^3$. We do this in such a way that the $U(1) \times SO(3)$
isometries of the geometry are preserved.
\label{fig:surgery2}
}
\end{figure}

The singularity that appears in the Ricci flow of the black hole is
quite analogous. Again, we have an $S^2$ fibered over a base, which in
this case is a disk. And again, the fiber over a point on the base
collapses to zero size after a finite flow time. We are therefore
inspired by the above example to perform a similar surgery in order to
continue the flow. Specifically, we excise a small disk from the base
in a neighborhood of the singular fiber. The base thus becomes an
annulus, and on its new inner boundary the fiber radius is made to go
to zero in order to produce a smooth total space with topology
$S^1\times B^3$; the $S^2$, which was non-contractible, is now
contractible, and vice-versa for the $S^1$. The change to the metric
is localized to a small distance $\epsilon$ from the singularity, and
is required to respect the geometries $U(1)\times SO(3)$ isometry
group. A cartoon of it is shown in figure \ref{fig:surgery2}. The
resulting geometry can be seen in the example of figure
\ref{fig:flowminus1} in the first snapshot of the bottom
panel. Details of how the surgery was carried out are given in
Appendix B.

Those who are interested in Ricci flow chiefly insofar as it
approximates the RG flow of sigma models might ask the following
question: What happens to the RG flow when the Ricci flow hits the
singularity? When the curvatures become large in string units, the
$\alpha'$ corrections to the beta functions \eqref{RG} become
important, and Ricci flow is no longer a good approximation to RG
flow. However, in general this will not be sufficient to stop a
singularity from forming. This is shown by a simple example, namely
the round metric on $S^D$ (otherwise known as the $O(N)$ model, with
$N=D+1$). On the Ricci flow side, the sphere collapses to zero size in
finite time. On the RG flow side, the model confines at a finite
energy scale, and is massive below that scale. In the case of a sphere
fibered over a base, it seems reasonable to guess that, similarly,
when the fiber over some point on the base collapses (``confines"), it
is removed from the target space, whose topology would change
accordingly.\footnote{This is similar to the picture of
\cite{Adams:2005rb}, but in that case the fiber is an $S^1$ and the
confinement is driven by the condensation of a winding tachyon.} In
other words, we are proposing that, when the Ricci flow hits the
singularity, the RG flow dynamically performs the surgery we have
described above. The fact that this surgery is the unique local
topology change that respects the manifold's isometries means that the
RG flow could hardly do otherwise.

After the surgery, the manifold has the same topology as hot flat
space, and under Ricci flow the metric relaxes to that saddle point
without encountering any further singularities. This can be seen in
the bottom panel of figure \ref{fig:flowminus1}; the same behavior was
observed for the other values of $b$ tested. A key assumption is that,
provided the metric deformation in the surgery is restricted to a
region of size $\epsilon$, for small enough $\epsilon$ the details of
the flow following the resolution will be independent of the precise
nature of this deformation, and of the actual value of
$\epsilon$. This is to be expected since the diffusive nature of Ricci
flow quickly removes the fine details of the small resolved
region. While we do not have a rigorous mathematical argument that
this is the case, we found it to be true in our simulations, as shown
in Appendix B. (The value of $\epsilon$ used to generate the
singularity resolution in figure \ref{fig:flowminus1} was $0.07$).

%
\section{A novel free energy diagram}
\label{sec:phase}
%

\begin{figure}
\centerline{\epsfig{file=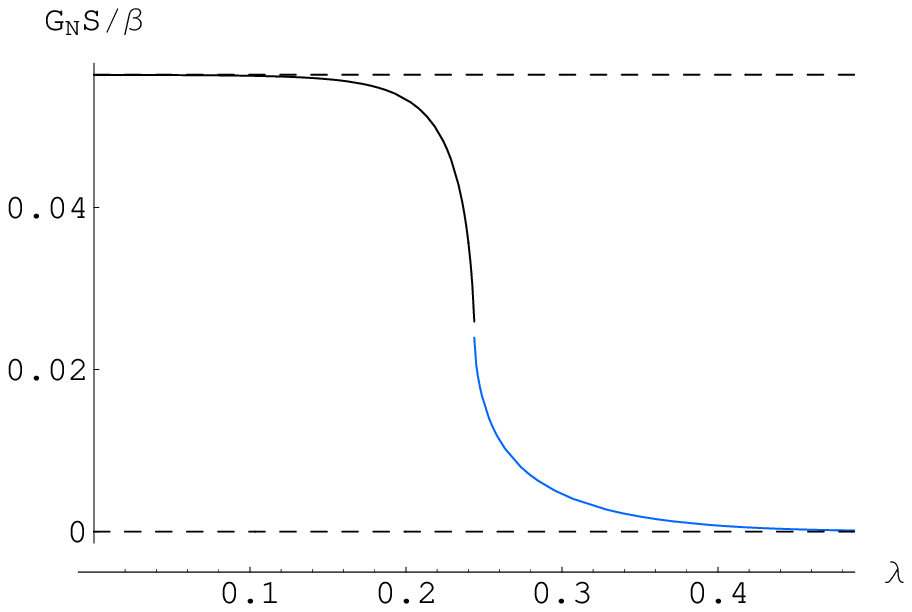,width=3.5in}\epsfig{file=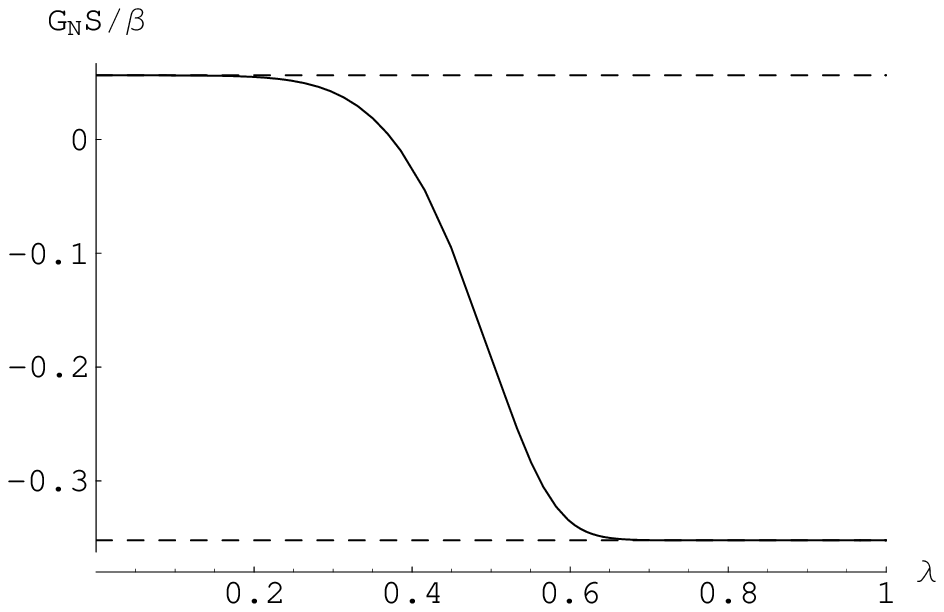,width=3.5in}}
\centerline{\epsfig{file=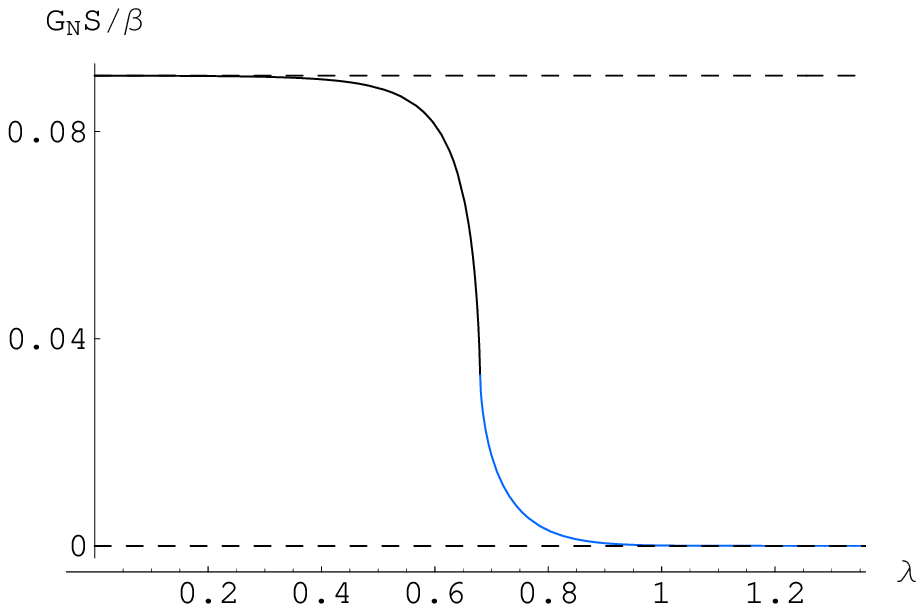,width=3.5in}\epsfig{file=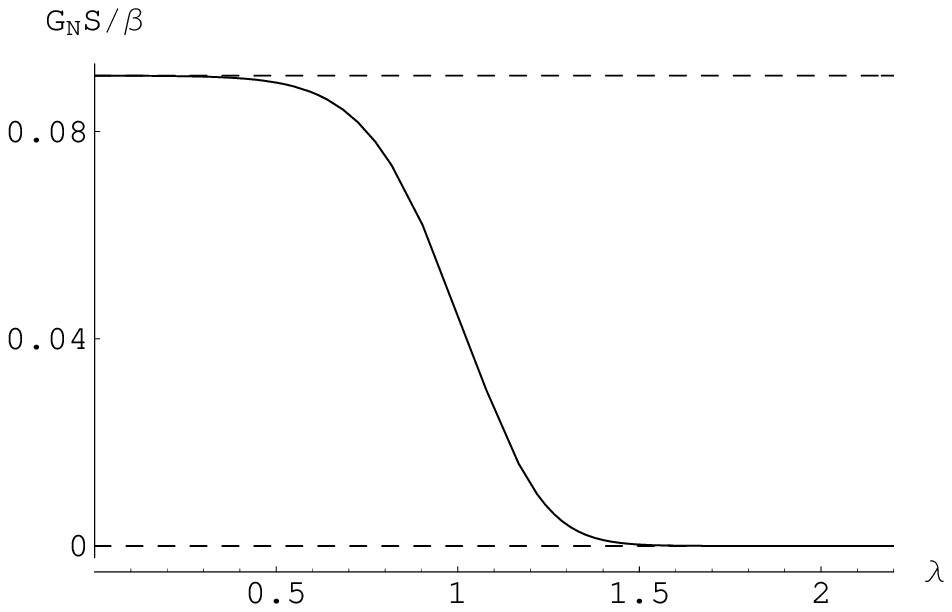,width=3.5in}}
\centerline{\epsfig{file=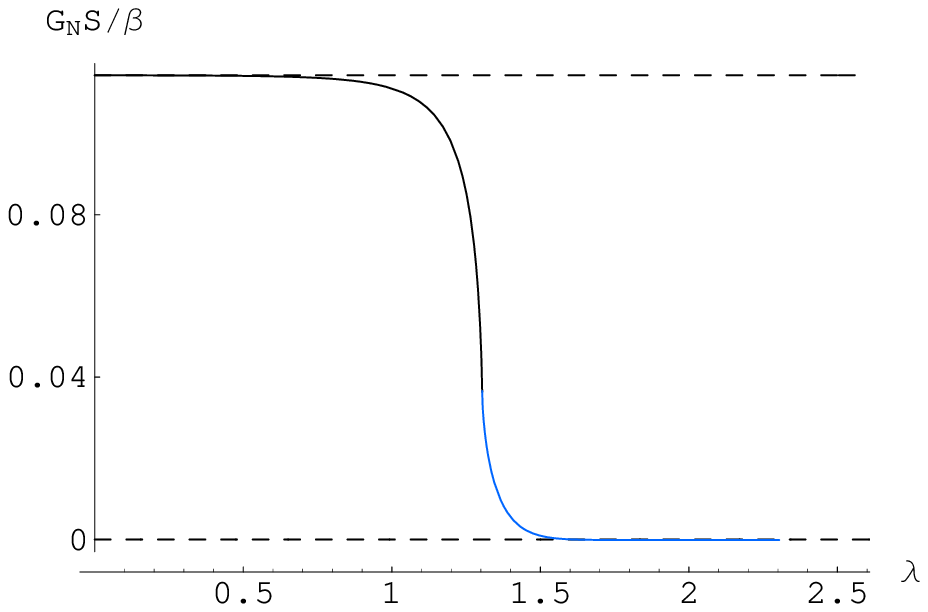,width=3.5in}\epsfig{file=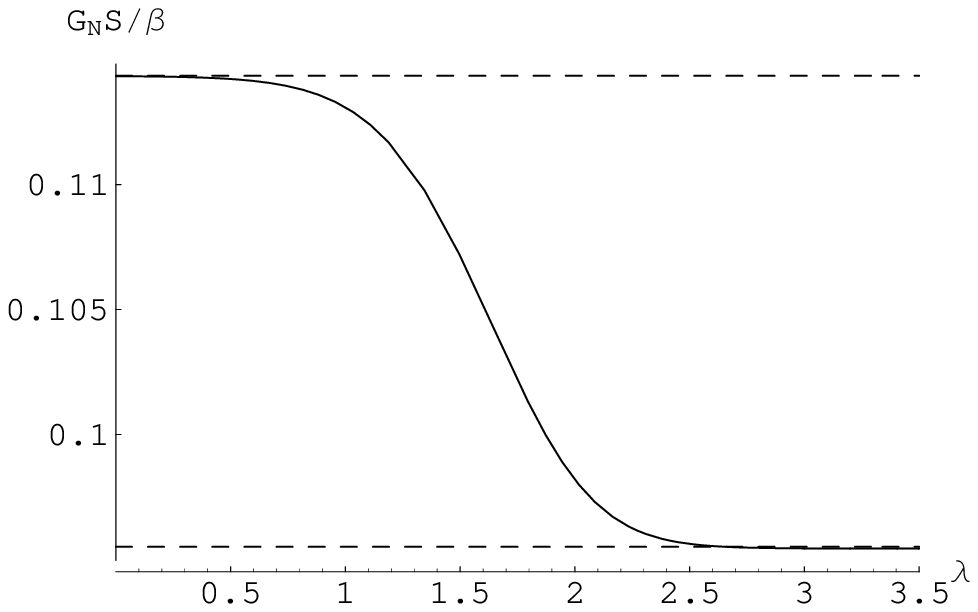,width=3.5in}}
\caption{Action $S$ versus flow time for the flows from the small
black hole to hot flat space (left) and the large black hole (right),
at $b=0.40,0.59,0.70$ (top, middle, bottom). The black lines indicate
the topology $B^2\times S^2$, and the blue the topology $S^1\times
B^3$, their join in the left plots being the point of singularity
formation where the action remains continuous (up to a small numerical
error in the case $b=0.40$).
\label{fig:freeerg}
}
\end{figure}

\begin{figure}
\centerline{\epsfig{file=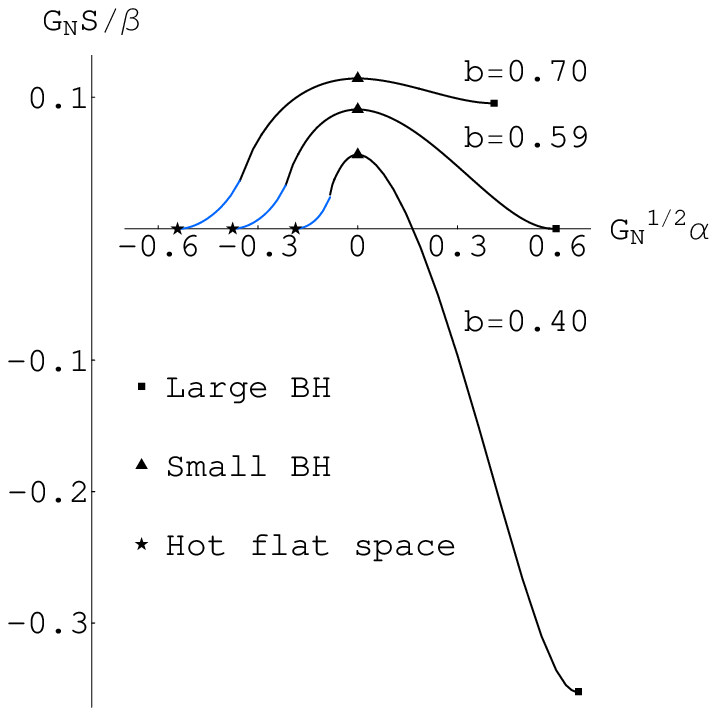,width=4in}}
\caption{$S / \beta$ against proper distance along the one-parameter
family in the space of metrics defined by Ricci flow.  In the
left-hand (blue) part of each curve the topology is $S^1\times B^3$,
while in the right-hand (black) part the topology is $B^2\times
S^2$. As explained in the text, for these curves the value of
$S/\beta$ can be interpreted as a free energy.
\label{fig:phase}
}
\end{figure}

In the last section we saw that Ricci flow defines paths in the space
of metrics connecting the small black hole to the large black hole and
to hot flat space, in the latter case via a topology change. Since
Ricci flow is gradient flow of the Einstein-Hilbert action
\eqref{eq:action}, it is interesting to plot the action along the
flow. This is done in figure \ref{fig:freeerg} for the two flows at
the three representative values of $b$ used in previous figures. We
would like to point out two interesting features of these
plots.

First, in the flow to hot flat space, the action stays finite when the
singularity forms and is continuous across the topology change;
although the Ricci scalar blows up on the horizon when it collapses,
it remains integrable. The local singularity model described in
Appendix B indicates that the divergence in the Ricci scalar is
entirely due to the shrinking 2-sphere, so that $R \sim 1/S^2$ at the
horizon. This is compensated by the factor of the 2-sphere volume in
the measure $\sqrt{g} \sim T S^2$, and since $T \rightarrow 0$ at the
horizon too, there is no contribution to the action from this singular
point. From the point of view of the action, therefore, the
singularity is harmless, and the surgery can be thought of as an
infinitesimal change in the metric.

Second, the action is a monotonically decreasing function of $\lambda$. This is \emph{not} a straightforward consequence of the fact that Ricci flow is gradient flow of the action, since the metric $G_{AB}$ is not positive definite (which is why we have  used the term gradient \emph{flow}, rather than gradient \emph{descent}). The monotonicity of $S$ for these particular flows shows that they move through the space of metrics always in a spacelike direction; in other words, at every point along the flow we have 
\begin{equation}
-\pd S\lambda = G_{AB}\pd{g^A}\lambda\pd{g^B}\lambda =
\frac1{8 \pi G_{\rm N}}\int_M\sqrt g\left(R_{\mu\nu}R^{\mu\nu}-\frac12R^2\right) > 0.
\end{equation}
Hence the proper distance along the curves is well defined; it is given by
\begin{equation}
\alpha \equiv \int d\lambda\sqrt{G_{AB}\pd{g^A}\lambda\pd{g^B}\lambda}.
\end{equation}
In figure \ref{fig:phase}, we plot $S$ against $\alpha$, putting the
two flows starting at the small black hole back to back where they
join smoothly. Note that the saddle points are a finite distance away
from each other along these curves.

Figure \ref{fig:phase} bears a strong resemblance to a free energy
diagram. Can it be interpreted as such in a physically meaningful way?
To clarify this question, let us first briefly review how free energy
diagrams are conventionally defined. Suppose one is interested in
knowing the expectation value of some macroscopic observable $a$ (for
example, the magnetization of a ferromagnet). To do this,
one calculates the free energy $f(a)$ in the sub-ensemble where the
value of $a$ is fixed; this calculation can, for example, be done in a
saddle point approximation. The expectation value of $a$, and the
actual free energy $F$ of the system, are determined by integrating
$e^{-\beta f(a)}$ over $a$, which, again in a saddle point
approximation, amounts to minimizing $f(a)$. In other words, in
calculating the partition function the degrees of freedom are
integrated over in two steps: first everything but $a$, then $a$
itself. The free energy diagram appears between these steps.

\begin{figure}
\flushleft{\epsfig{file=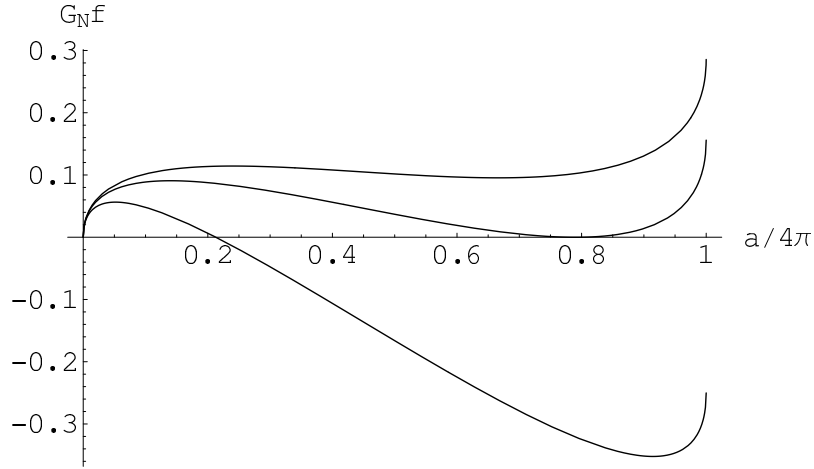,width=5.5in}}
\caption{Conventional free energy diagram for gravity in a cavity, in which the free energy $f$ is calculated in sub-ensembles where the horizon area $a$ is fixed. In the saddle point approximation, $f=S/\beta$, where $S$ is the action of the Schwarzschild metric with horizon area $a$ and a conical singularity in the transverse directions. The three curves are for the same values of $b$ as in the previous figure.
\label{fig:York}
}
\end{figure}

Let us illustrate these ideas for the case at hand, gravity in a cavity; our discussion follows York \cite{York:1986it}. We choose as our macroscopic observable $a$ the horizon area. Working on the topology $B^2\times S^2$, the horizon area is fixed by adding a Lagrange multiplier term to the action:
\begin{equation}
S_a[g] = S[g] + \lambda\left(A[g]-a\right).
\end{equation}
This term introduces into the Einstein equation a source term in the form of a two-dimensional delta function on the disk at the location of the horizon, leading to a conical singularity. The saddle point of the new action is therefore the Euclidean Schwarzschild metric \eqref{eq:metric}, but with $r_0$ determined by $4\pi r_0^2=a$ rather than by \eqref{eq:betaeq}, and with the periodicity of $\tau$ equal to $\beta/(2r_0\sqrt{1-r_0/R})$ rather than $2\pi$. The action of this solution is easily calculated, giving a free energy of the familiar form ``energy minus temperature times entropy":
\begin{equation}
f = \frac S\beta = 
\frac R{G_{\rm N}}
\left[1-\sqrt{1-\sqrt{\frac a{4\pi R^2}}}-\frac a{4\beta R}\right].
\end{equation}
Following Whiting and York \cite{WhitingYork}, we plot $f$ in figure
\ref{fig:York}, where the three saddle points are easily
identified. Hot flat space is at $a=0$; in fact the topology change
from $B^2\times S^2$ to $S^1\times B^3$ occurs at that very point on
the free energy diagram (which to some extent excuses the rather
singular behavior of $f$ there).

Let us now return to our putative free energy plot, figure
\ref{fig:phase}, derived from the Ricci flows. $S/\beta$, which is plotted
for each point $g^A(\lambda)$ along the flow, can justifiably be
called a free energy if it is the saddle point approximation to the
partition function of some sub-ensemble. We will now show that it is. First note that the gradient of $S$, restricted to the
directions orthogonal to the curve $g^A(\lambda)$, vanishes, since the
tangent vector to the curve is in the gradient direction. Furthermore
the second-derivative operator ${\Delta^A}_C$ (defined in
\eqref{Delta}) restricted to those directions is positive, since its
only negative direction is along the curve. Therefore the saddle point
approximation to the path integral along the directions orthogonal to
the curve is given by the value of the action on the curve. This
defines the sub-ensembles for which figure \ref{fig:phase} is the
appropriate free energy diagram. It is obviously crucial for this
construction that $g^A(\lambda)$ is defined by a gradient flow of the
action. In fact the same construction could be applied to any system
equipped with a natural metric on the configuration space. The major
difference between this construction and the conventional one is that
this one does not require singling out a particular macroscopic
observable.

We note that York's prescription of fixing the horizon area is
motivated by the Lorentzian dynamics of an evaporating black hole,
since the horizon size is the slow variable in\ that process.  From
the point of view of the canonical Euclidean path integral, York's
construction is less natural, as seen by the conical singularity that
arises in the Euclidean geometries.

We conclude by commenting that in the context of the AdS-CFT duality
it is interesting to consider the constrained free energy of
asymptotically AdS spacetimes with and without horizons. On the field
theory side of the correspondence, at weak coupling, one can naturally
compute such an off-shell free energy \cite{Minwalla}, which has a
similar appearance to ours. At strong coupling, this is expected to be
similar, with the extrema corresponding on the gravity side to exactly
the small and large AdS-Schwarzschild black holes, and thermal AdS,
and the Hawking-Page phase transition being interpreted in field
theory as the confinement/deconfinement transition. We note however,
that the constraint in the field theory is placed on the expectation
value of the trace of the Polyakov loop. On the gravity side, this
presumably translates into a constraint on the area of a string world
sheet wrapping the thermal Euclidean time circle at the asymptotic AdS
boundary. Hence this is a different constraint than the one imposed by
York; it would be interesting to find the relevant constrained
off-shell geometries.\footnote{See \cite{Barbon:2001di,Barbon:2002nw,Barbon:2004dd,Myung} for related work on this system.}

%
\section*{Acknowledgments}
%

We are grateful to G. Gibbons, S. Hartnoll, G. Horowitz, L. Lehner,
R. McNees, S. Minwalla, M. Perry, J. Sparks and D. Tong for useful
conversations, and to S. Minwalla for useful comments on the
manuscript. We would also like to thank the Perimeter Institute for
Theoretical Physics, the Kavli Institute for Theoretical Physics, and
Cambridge University for hospitality while this work was being
completed. The research of M.H. is supported by a Pappalardo
Fellowship and by the U.S. Department of Energy through cooperative
research agreement DF-FC02-94ER40818. The research of T.W. is
supported by NSF grant PHY-0244821.

\appendix

%
\section{Details of Ricci flow simulation}
\label{app:method}
%

In practice we choose units so that $R = 1$. Then the small black
holes are given by \eqref{eq:metric} for $r_0 < \frac{2}{3}$, and the
large black holes with $\frac{2}{3} < r_0 < 1$. Our boundary
conditions on the box wall are fixing the sphere radius ($R = 1$), and
fixing the Euclidean time period $\beta = 2 \pi b$. We start with the
small black hole metric, perturbed by a suitably small
$h^{\mathrm{GPY}}$ which satisfies the above boundary
conditions. For the data presented here, the amplitude of the
perturbation was chosen to change the horizon radius by one
percent. The initial perturbation $\pm h^{\mathrm{GPY}}$ respects
the $U(1) \times SO(3)$ isometry of the black hole, and hence the
entire flow will share this isometry. We exploit this, using the
metric,
\begin{equation}
ds^2 = e^{2 A} r^2 b^2 d\tau^2 + e^{2 B} dr^2 + e^{2 C} d\Omega_2^2
\label{eq:horizmetric}
\end{equation}
where $\tau$ has period $2 \pi$. Since we initially are interested in
the topology $B^2 \times S^2$ then $A,B,C$ are finite functions of $r$
and the flow parameter $\lambda$, and we take $0 \leq r \leq 1$. At $r
= 1$ we take $A = 0, C = 0$ appropriate to our finite box boundary
condition. The new `boundary' at $r = 0$ is fictitious, since we
know the Euclidean black hole geometries are smooth manifolds without
an interior boundary. Regularity of the Ricci flow equations implies
$A_{,r} = B_{,r} = C_{,r} = 0$ at this origin. In particular, one
finds as expected that $(A/C)|_{r = 0}$ is preserved under the flow,
and hence an initially smooth 4-d solution remains smooth, without
developing conical singularities.

For asymptotically flat spacetime we are free to make any choice of
$\xi_{\mu}$ in \eqref{diffeoRF}. However, on a manifold with boundary
we must take care. We have covered our manifold with coordinates such
that the boundary is at $r = 1$. Hence Ricci flow with $\xi_{\mu} = 0$
is only equivalent to flow with non-vanishing $\xi_{\mu}$ provided
$\xi_{r} = 0$ at $r = 1$. Otherwise the coordinate location of the
boundary will change during the flow.

Subject to $\xi_r = 0$ at the box wall, we are free to specify any
$\xi_{\mu}$ in \eqref{diffeoRF}. We make the choice,
\begin{equation}
\xi^{\mu} = - \Delta_{\rm S} x^{\mu} + \Delta_{\rm S} x^{\mu} |_{A,B,C = 0} 
\end{equation}
where $x^{\mu}$ are the coordinates and $\Delta_{\rm S}$ is the scalar
Laplacian. This choice is similar to the DeTurck flow, and ensures
strong parabolicity---in analogy with the harmonic coordinate choice
which yields strongly elliptic Euclidean Einstein equations. The flow
equation becomes,
\begin{equation}
\frac{d}{d \lambda} g_{\mu\nu} = \Delta_{\rm S} g_{\mu\nu} + F( g, \nabla g)   
\end{equation}
where on the right-hand side we have only explicitly displayed the second
derivative terms, which are simply given by the scalar Laplacian. This
parabolic flow requires boundary conditions for all components of the
metric, namely $A,B,C$, at the cavity wall. There we take $A = 0$, $C
= 0$, since we fix the induced metric there. The only non-zero
component of $\xi_{\mu}$ is,
\begin{equation}
\xi_r = - \partial_r A + \partial_r B - 2 \partial_r C
\label{eq:gauge}
\end{equation}
and hence we see the boundary condition that $\xi_r = 0$ at $r = 1$
translates into a boundary condition of the gradient of $B$. Hence
$A,B,C$ all have boundary conditions on their value, or gradient,
compatible with the parabolic flow.

It is interesting to note that the 4-d Schwarzschild metric takes the
elegant analytic form,
\begin{equation}
ds^2 = r_0^2 \left[ 4 (1-r_0) \left( r^2 d\tau^2 + \frac{1}{( 1 - (1-r_0) r^2 )^4} dr^2 \right) + \frac{1}{( 1 - (1-r_0) r^2 )^2} d\Omega_{2}^2 \right]
\label{eq:altmetric}
\end{equation}
in the coordinate system where $\xi^{\mu} = 0$ everywhere, and furthermore the
derived metric functions $A,B,C$ are finite as we require. Hence we
take this as a background metric. We compute $\pm h^{\mathrm{GPY}}$
using the Lichnerowicz operator in this background, also in the gauge
$\xi^{\mu} = 0$. As usual, the resulting coupled ordinary differential
equations are best solved as a shooting problem. Note that the
condition $\xi_r = 0$ together with $A = C = 0$ implies that we cannot
expect $B = 0$ at the cavity wall. As mentioned in the main text, this
means the mode cannot be traceless there, which clearly would also
require $B = 0$. Hence the cavity wall couples the trace and traceless
perturbations, unlike in the asymptotically flat or AdS computations.

Having computed this, we use it to perturb the small
black hole with the amplitude chosen as described above. This forms
our initial data for the Ricci flow, and ensures our condition $\xi_r
= 0$ is satisfied at the box wall.

We simulated the full Ricci flows for
\begin{equation}
b = 0.10, 0.20, 0.30, 0.40, 0.59, 0.62, 0.64, 0.66, 0.68, 0.70, 0.72, 0.74.
\end{equation}
The value $b=16/27\approx 0.59$ gives the middle dotted line in figure
\ref{fig:box} where the free energies of hot flat space and the large
black hole are equal. The values $0.40$ and $0.70$ give the other
dotted lines in this figure. Properties of the flows for these three
values are displayed in the various figures here, but we obtained
qualitatively similar behavior for all values of $b$ above.

We used a uniform lattice discretization, and solve the Ricci flow
using the implicit Crank-Nicholson method since it has diffusive
character. We performed flows at various resolutions and confirmed the
correct convergence to the continuum. Data presented here use $dr =
0.0025$, i.e.\ 400 points.

After singularity resolution for the $- h^{\mathrm{GPY}}$ flows, we
are interested in the topology $S^1 \times B^3$ where we use the
metric,
\begin{equation}
ds^2 = e^{2 \tilde{A}} b^2 d\tau^2 + e^{2 \tilde{B}} dr^2 + e^{2
\tilde{C}} r^2 d\Omega^2
\label{eq:flatmetric}
\end{equation}
and take finite $\tilde A,\tilde B,\tilde C$, again with $0 \leq r
\leq 1$. Now regularity of the Ricci flow equations at $r = 0$
requires the boundary conditions $\tilde B = \tilde C$ and $\tilde
A_{,r} = \tilde B_{,r} = \tilde C_{,r} = 0$ at the origin, ensuring
again a regular 4-d Euclidean geometry. As for \eqref{eq:horizmetric}
we make the choice,
\begin{equation}
\xi^{\mu} = - \Delta_{\rm S} x^{\mu} + \Delta_{\rm S} x^{\mu} |_{\tilde{A},\tilde{B},\tilde{C} = 0},
\end{equation}
again ensuring a strongly parabolic flow. As above, we require $A = C
= 0$ at the cavity wall, and obtain a boundary condition for $B$ from
the condition that $\xi_r$ vanishes there.

%
\section{More on the singularity and surgery}
\label{app:sing}
%

\begin{figure}
\centerline{\epsfig{file=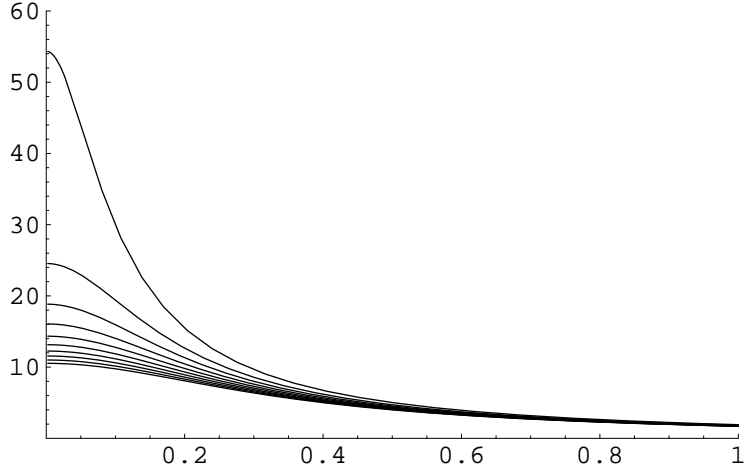,width=4in}}
\caption{$(R_{\alpha\beta\mu\nu}R^{\alpha\beta\mu\nu})^{1/4}$ against $r$, at different values of flow time as the singularity is approached, for the flow seeded by $-h_{\mu\nu}^{\rm GPY}$ with $b = 0.59$.
\label{fig:kretch}
}
\end{figure}

\begin{figure}
\centerline{\epsfig{file=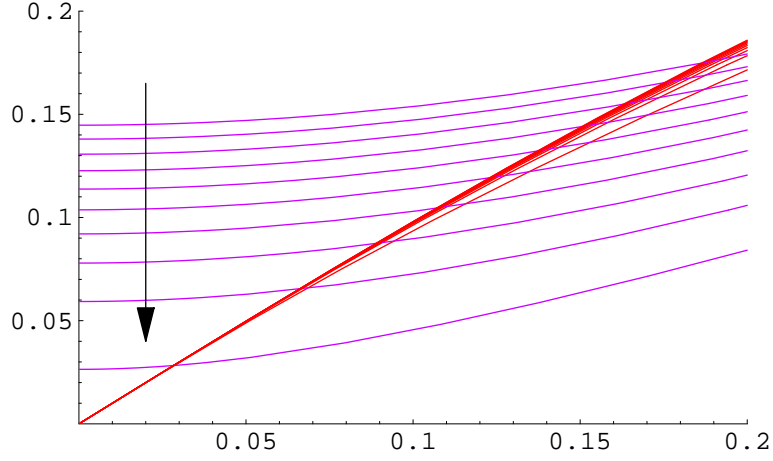,width=4in}}
\caption{$T(\rho)$ and $S(\rho)$ plotted against proper distance from
the horizon, $(\rho-\rho_H)$, in the vicinity of the horizon as the
singularity is approached. Snapshots are taken at intervals of
$\lambda$ of $0.002$.
\label{fig:sing}
}
\end{figure}

In figure \ref{fig:kretch} we plot the curvature invariant
$(R_{\alpha\beta\mu\nu}R^{\alpha\beta\mu\nu})^{1/4}$ for the flow
generated by $-h^{\rm{GPY}}_{\mu\nu}$. We see that, as expected, the
shrinking of the 2-sphere at the horizon leads to an increasingly
singular curvature there. This nature of the singularity that forms
may be seen by plotting $T,S$ in the vicinity of the horizon, as in
figure \ref{fig:sing}. We see that since the geometry is smooth before
the singular flow time, the gradient of $T$ is fixed at the
horizon. The interesting part of the geometry is then the shrinking
2-sphere, and we see that whilst the size of this at the horizon
decreases towards the singularity, the rate of expansion of the sphere
away from the horizon appears to remain zero as the singularity is
approached. Hence the local model for this singularity is simply
$\mathbf{R}^2 \times S^2$. Note that the only other option we know of
for the singularity model would have been a cone with base $S^1 \times
S^2$.

\begin{figure}
\centerline{\epsfig{file=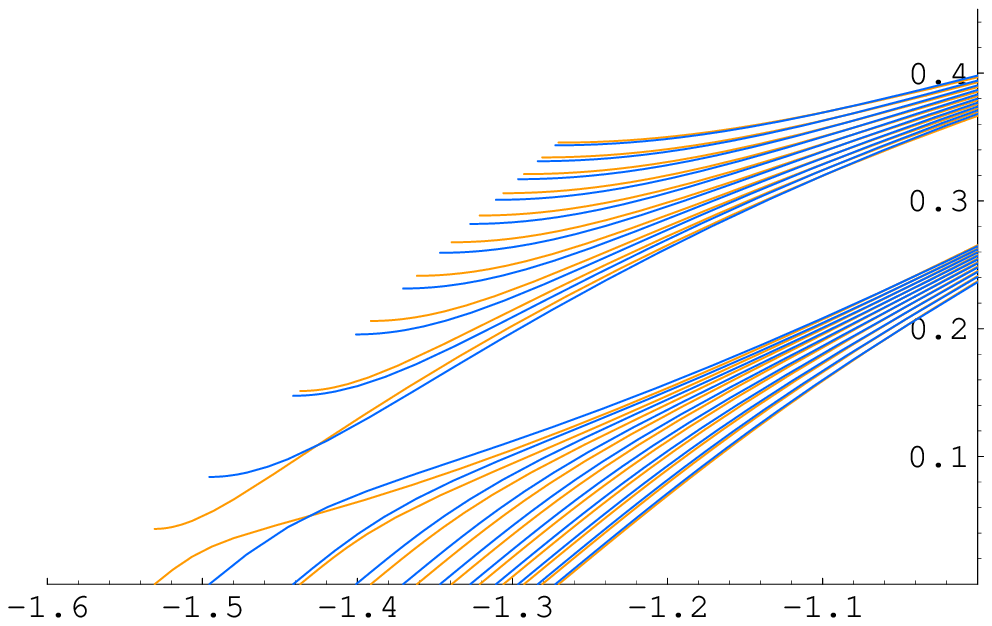,width=4.5in}}
\caption{Comparison of two different resolutions of the singularity
formed in the $b = 0.59$ flow. The size of the region where the
surgery is performed is different in the two cases, and characterized
by $\epsilon = 0.07$ (orange) and $0.14$ (blue), but the resulting
flows appear very similar after a short time. The geometries are
plotted for intervals $0.002$ in $\lambda$ after the resolution. Both
flows eventually tend to flat space, although this is not shown here
for clarity.
\label{fig:flowminus3}
}
\end{figure}

We wish to perform surgery on the geometry just before the singularity
is reached. We do this by taking the geometry at some flow time
$\lambda_{0}$ before the singularity, with functions $A, B, C$ in
\ref{eq:horizmetric}, and moving to the topology \ref{eq:flatmetric},
with,
\begin{eqnarray}
e^{2 \tilde{A}} & = & (\epsilon^2+r^2) e^{2 A} \\
e^{2 \tilde{B}} & = & e^{2 B} \\
e^{2 \tilde{C}} & = & \frac{1}{(\epsilon^2+r^2)} e^{2 C} 
\end{eqnarray}
with,
\begin{equation}
\epsilon = e^{C-B} |_{r=0}.
\end{equation}
Note the resulting geometry with topology $S^1 \times B^3$ is smooth,
and for $r\gg\epsilon$ the geometry before and after surgery is the
same, the modification only occurring for $r \sim \epsilon$. The size
of $\epsilon$ therefore determines the size of the region where
surgery is performed. This is smaller the closer one takes
$\lambda_{0}$ to the singularity. In figure \ref{fig:flowminus3} we
show two flows for $b = 0.59$ which use different values of
$\lambda_{0}$, giving $\epsilon = 0.07$ and $0.14$. We note that both
do indeed flow to flat space, and furthermore, the large scale form of
the flow appears to be insensitive to the details of the resolution.


%
\bibliography{ref}
\bibliographystyle{JHEP}
%

%
\end{document}